\documentclass[11pt,english,a4paper,floatfix]{article}
\pdfoutput=1
\usepackage{jcappub}

\usepackage{bm}
\usepackage{amsfonts}
\usepackage{subfig,graphicx}
\usepackage{url}
\usepackage{color}
\usepackage[dvipsnames]{xcolor}
\usepackage[normalem]{ulem}

\newcommand{\alt}{\mbox{\;\raisebox{.3ex}
  {$<$}$\!\!\!\!\!$\raisebox{-.9ex}{$\sim$}\;}}

\newcommand{\be}{\begin{equation}}
\newcommand{\ee}{\end{equation}}
\newcommand{\bea}{\begin{eqnarray}}
\newcommand{\eea}{\end{eqnarray}}

\usepackage{lmodern}
\usepackage{slantsc}

\newcommand{\appropto}{\mathrel{\vcenter{
  \offinterlineskip\halign{\hfil$##$\cr
    \propto\cr\noalign{\kern2pt}\sim\cr\noalign{\kern-2pt}}}}}

\usepackage[scr=boondox]{mathalfa}

\usepackage{mathtools}

\begin{document}


\title{What will it take to measure individual neutrino mass states using cosmology?}

\author[a,b]{Maria Archidiacono}

\affiliation[a]{Universit\'a degli Studi di Milano, via G. Celoria 16, 20133 Milano, Italy}

\affiliation[b]{INFN Sezione di Milano, via G. Celoria 16, 20133 Milano, Italy}

\author[c]{, Steen Hannestad}

\affiliation[c]{Department of Physics and Astronomy, Aarhus University,
 DK-8000 Aarhus C, Denmark}

\author[d]{, Julien Lesgourgues}
\affiliation[d]{Institute for Theoretical Particle Physics and Cosmology (TTK), \\ RWTH Aachen University, D-52056 Aachen, Germany}

\emailAdd{maria.archidiacono@unimi.it}\emailAdd{sth@phys.au.dk}\emailAdd{lesgourg@physik.rwth-aachen.de}

\abstract{We study the impact of assumptions made about the neutrino mass ordering on cosmological parameter estimation with the purpose of understanding whether in the future it will be possible to infer the specific neutrino mass distribution from cosmological data. We find that although the commonly used assumption of a degenerate neutrino hierarchy is manifestly wrong and leads to changes in cosmological observables such as the cosmic microwave background and large scale structure compared to the correct (normal or inverted) neutrino hierarchy, the induced changes are so small that even with extremely optimistic assumptions about future data they will remain undetectable. We are thus able to conclude that while cosmology can probe the neutrino contribution to the cosmic energy density extremely precisely (and hence provide a detection of a non-zero total neutrino mass at high significance), it will not be possible to directly measure the individual neutrino masses.
}

\hfill{\small TTK-20-05}

\maketitle


\section{Introduction}

Over the past decade cosmology has proven to be an extremely powerful tool for probing neutrino physics. 
For example, using a variety of different data, primarily from the cosmic microwave background (CMB) and large scale structure (LSS), it has been possible to firmly constrain the neutrino contribution to the cosmic energy density (see, e.g., Refs.\ \cite{Aghanim:2018eyx,Palanque-Delabrouille:2019iyz,RoyChoudhury:2019hls}).

The main effect of sub-eV mass neutrinos on late time structure formation is to suppress growth on all scales inside their particle horizon, which is the distance over which they can travel since their creation in the early universe \cite{Bond:1980ha}. While this effect can be probed using current structure formation data \cite{Hu:1997mj}, it is necessary to add early-time information from the CMB in order to separate the neutrino contribution from other effects \cite{Archidiacono:2016lnv}.
Using this data combination it has been possible to constrain the contribution from a neutrino-like component to be $\Omega_\nu h^2 \alt 0.001$, depending somewhat on the specific data sets used \cite{Aghanim:2018eyx,Palanque-Delabrouille:2019iyz,RoyChoudhury:2019hls}. Assuming standard model neutrino physics this can be translated into an upper bound on the sum of all neutrino masses from states that are currently non-relativistic \cite{Hannestad:2006zg,Lesgourgues:2006nd,Wong:2011ip,Lesgourgues:1519137} of $\sum m_\nu \alt 0.1$ eV.

However, this constraint depends little on the exact details of neutrino physics. In reality, the bound applies to any component which is weakly interacting, relativistic around the CMB last scattering surface, and non-relativistic at the current epoch. This means that the bound can be applied to other particles such as axions \cite{Archidiacono:2013cha,DiLuzio:2020wdo}. However, it also means that cosmology is not a strong probe of neutrino properties beyond their contribution to the background and perturbed stress-energy tensor: it does not significantly constrain the neutrino phase-space distribution (see, e.g., Refs.\ \cite{Cuoco:2005qr,deSalas:2018idd,Oldengott:2019lke}) and simply excludes the possibility of too strong non-standard neutrino interactions in cases where particle physics experiments do not already provide stronger bounds (see, e.g., Refs.\ \cite{Archidiacono:2013dua,Wilkinson:2014ksa,Oldengott:2017fhy,Diacoumis:2018ezi,Park:2019ibn,Blinov:2019gcj,Grohs:2020xxd}).

In cosmological parameter estimation a common assumption about the neutrino sector is that the three neutrino states have equal mass. While this assumption is manifestly wrong given our knowledge of neutrino mass splittings from oscillation experiments, it is sufficiently accurate for current parameter estimation purposes \cite{Lesgourgues:2004ps,Oyama:2012tq,Oyama:2015gma,DiValentino:2016foa,Hannestad:2016fog,Gerbino:2016ehw,Giusarma:2016phn,Schwetz:2017fey,Vagnozzi:2017ovm,Vagnozzi:2018jhn,Gariazzo:2018pei}.
Note that this conclusion would not apply to the even cruder approximation consisting in replacing the normal hierarchy by one massive and two massless neutrinos, or the inverted hierarchy by two massive and one massless neutrinos: the latter approximations lead to observables significantly different from the true ones, as pointed out in several papers since Ref.\ \cite{Lesgourgues:2004ps} (figure 4).

As mentioned above, when the current bound on $\Omega_\nu h^2$ is translated into a bound on the neutrino mass, current data give a bound of $\sum m_\nu \alt 0.1$ eV, where the sum runs over all currently non-relativistic states.  

Recent global analyses of oscillation data, such as NuFit 4.0 \cite{Esteban:2018azc}, provide the following best-fit values for the standard model neutrino mass splittings:
\begin{eqnarray} 
 \Delta m_{21}^2 & = & \left(7.39^{+0.21}_{-0.20}\right)\times 10^{-5} \, {\rm eV}^2, \nonumber \\
  \Delta m_{31}^2 & = & \left(2.525^{+0.033}_{-0.032}\right)\times 10^{-3} \, {\rm eV}^2 (\textrm{NH}), \label{eq:split} \\
  \Delta m_{32}^2 & = & \left(- 2.512^{+0.034}_{-0.032}\right)\times 10^{-3} \, {\rm eV}^2 (\textrm{IH}). \nonumber
\end{eqnarray}
Here, the value for $\Delta m_{21}^2$ is applicable to both the normal hierarchy (NH, $m_3>m_2>m_1$) and the inverted hierarchy (IH, $m_2 > m_1 > m_3$), while the other values are for NH and IH, respectively.

\begin{figure}[tbp]
\centering 
\includegraphics[width=.55\linewidth]{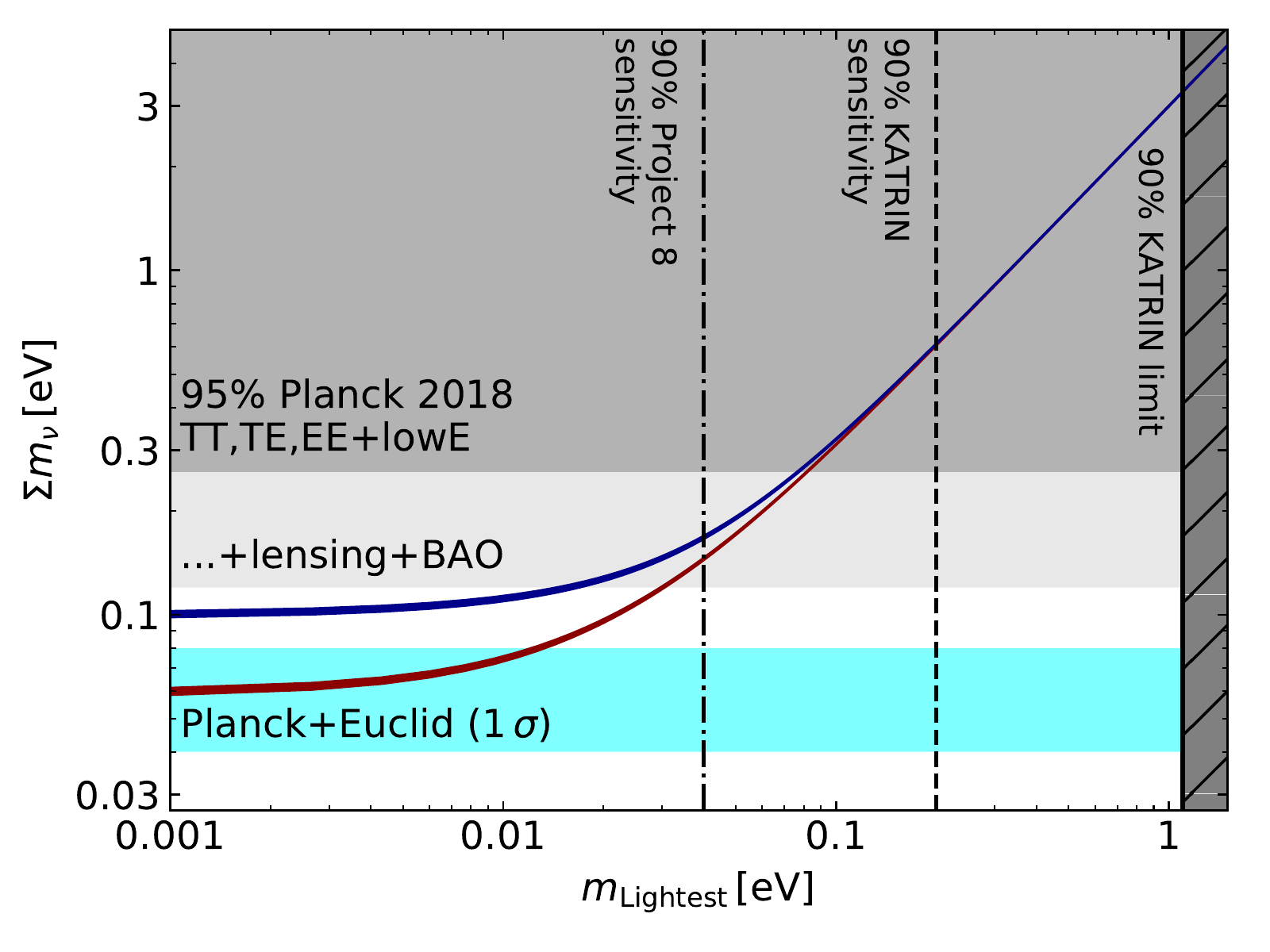}
\hfill
\caption{\label{fig:hierarchy}The neutrino mass sum as a function of the lightest neutrino mass for the inverted hierarchy (blue) and for the normal hierarchy (red).
The grey horizontal regions are currently disfavored at the 95\%CL by Planck 2018 TT,TE,EE+lowE (dim grey) and by Planck 2018 TT,TE,EE+lowE+lensing+BAO (light grey) within a minimal 7-parameter cosmology and assuming three neutrinos with the same mass \cite{Aghanim:2018eyx}.
The hatched grey vertical area is excluded by the recent KATRIN limit \cite{Aker:2019uuj}, assuming $m_\beta=m_\mathrm{Lightest}$, as it still lies in the degenerate region. The black dashed and the black dot-dashed vertical lines represent the expected sensitivity to the effective electron neutrino mass of, respectively, KATRIN and Project 8~\cite{Esfahani:2017dmu}.
The cyan band depicts the sensitivity of Planck+Euclid to a neutrino mass sum of $0.06$ eV ($\sigma(\sum m_\nu)=0.02$ eV), as obtained in the Markov Chain Monte Carlo forecast of Ref.\ \cite{Sprenger:2018tdb}, assuming three neutrinos with the same mass. Notice that the Euclid sensitivity can be further improved by including intensity mapping from future 21 cm radio telescopes such as the Square Kilometre Array \cite{Sprenger:2018tdb,Obuljen:2017jiy,Brinckmann:2018owf} or by adding higher order statistics \cite{Hahn:2019zob,Chudaykin:2019ock}.
}
\end{figure}

As shown in Fig.~\ref{fig:hierarchy}, the upper bound from cosmology is close to the minimum allowed by the inverted hierarchy, and therefore cosmology, when combined with oscillation data, shows a mild preference for the normal hierarchy over the inverted hierarchy
\cite{RoyChoudhury:2019hls} (see however Ref.\ \cite{Simpson:2017qvj}). However, this is almost entirely due to prior volume effects, i.e. the inverted hierarchy cuts away a significant portion of the preferred parameter space for cosmology.
Current cosmological data do not have the precision to probe the exact neutrino mass distribution \cite{Mahony:2019fyb}, and any information on the neutrino mass splittings and hierarchy must come from a combination with other (oscillation) data \cite{Gariazzo:2018pei,deSalas:2018bym}.

However, a question of significant interest is whether cosmology can {\it ever} achieve the level of precision necessary to probe the exact neutrino hierarchy, and a more detailed investigation of this question is indeed the purpose of this paper.
The question has previously been addressed in, e.g., Refs.\ \cite{Lesgourgues:2004ps,Takada:2005si,Pritchard:2008wy,Jimenez:2010ev,Hall:2012kg,Wagner:2012sw,Blennow:2013kga,Jimenez:2016ckl,Hannestad:2016fog,DiValentino:2016foa,Schwetz:2017fey}, but here we want to study it focussing on the underlying physics, and using full Markov Chain Monte Carlo parameter estimation techniques on synthetic data sets both with updated specifications about forthcoming surveys, and for futuristic surveys.

In Section \ref{sec:formalism} we will outline the physical differences induced by the exact neutrino mass distribution, both at the background and at the perturbation level, as well as in the cosmological observables. We will then in Section \ref{sec:results} proceed to look into forecasts using a variety of different cosmological probes, either forthcoming or suggested for the longer term future. In Section \ref{sec:conclusions} we provide a discussion and our conclusions.

\section{Physical effects of the neutrino mass splitting}
\label{sec:formalism}

The presence of neutrinos influences both the background expansion rate and the growth of perturbations. In order to assess the impact made by the choice of neutrino mass ordering we shall consider the following setup:
We assume that for NH $\sum m_\nu = 0.06$ eV, while for IH $\sum m_\nu = 0.10$ eV, i.e. values close to the minimal ones of each hierarchy.
The resulting mass distribution are:
\begin{align}
\mathrm{NH:} & \left(m_1=0.001075, m_2=0.008663, m_3=0.050261\right)\,\mathrm{eV}\\
\mathrm{IH:} & \left(m_1=0.049379, m_2=0.050122, m_3=0.000498\right)\,\mathrm{eV}.
\end{align}
We compare each hierarchy with its own degenerate case, i.e. DH implies $\sum m_\nu = 3 \times 0.02$ ($m_D \equiv \sum m_\nu /3 = 0.02$ eV) for NH and $\sum m_\nu = 3 \times 0.0333$ ($m_D=0.0333$ eV) for IH. We keep the other cosmological parameters ($\omega_b$, $\omega_{\rm cdm}$, $H_0$, $n_s$, $A_s$, $\tau_{\rm reio}$) fixed. Notice that here we keep the Hubble constant $H_0$ fixed, rather than fixing the angular scale of the sound horizon at recombination $\theta_s$, because we want to investigate the impact on the low redshift background probes independently on the CMB constraints. Later in Section \ref{sec:results} the combined forecast of CMB + LSS will use $\theta_s$ as free parameter, rather than $H_0$.

\subsection{Changes to background quantities}

\begin{figure}[tbp]
\centering 
\subfloat[]{\label{fig:bkg:nu}
\includegraphics[width=.45\linewidth]{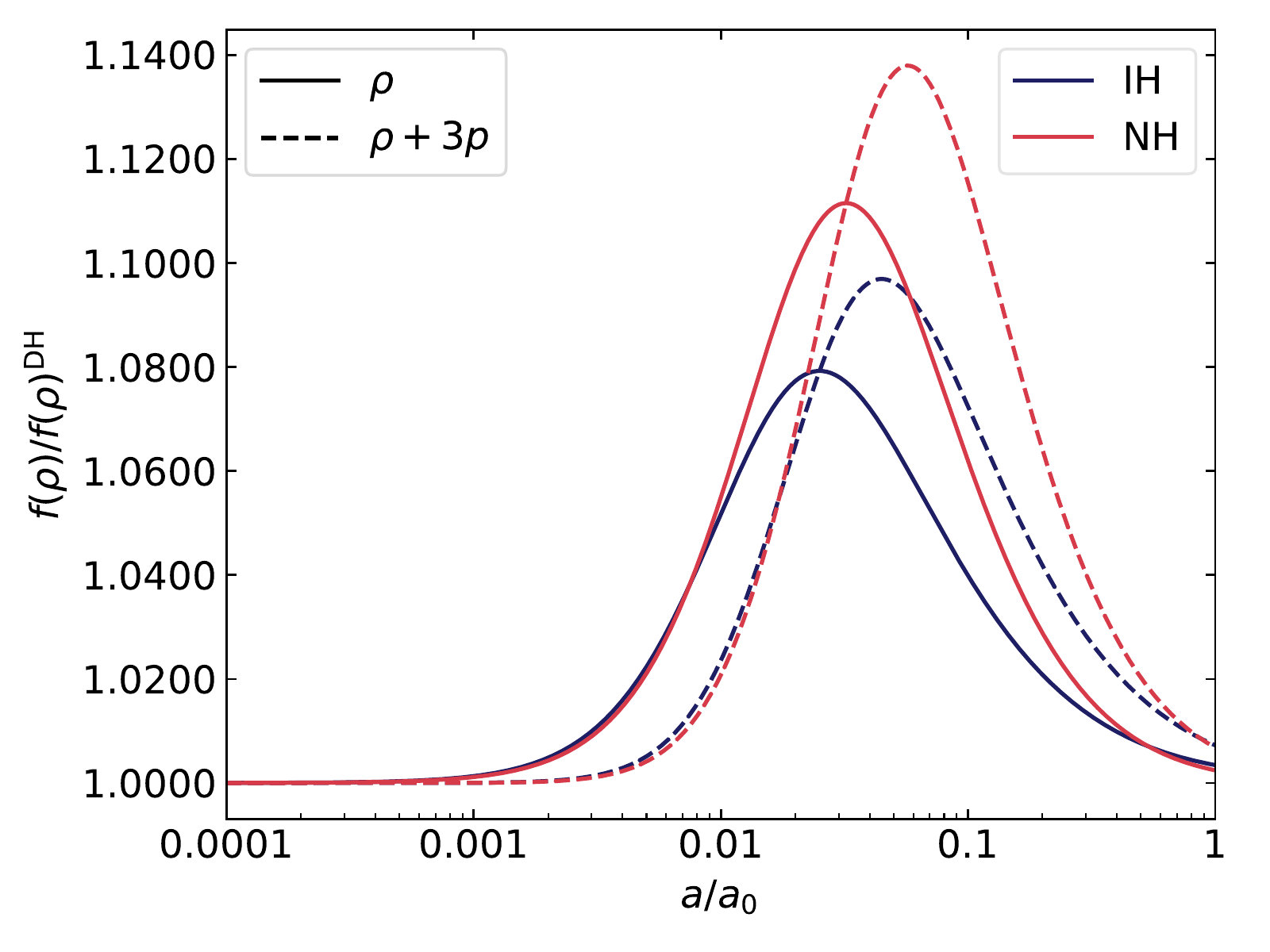}}\hfill
\subfloat[]{\label{fig:bkg:tot}
\includegraphics[width=.45\linewidth]{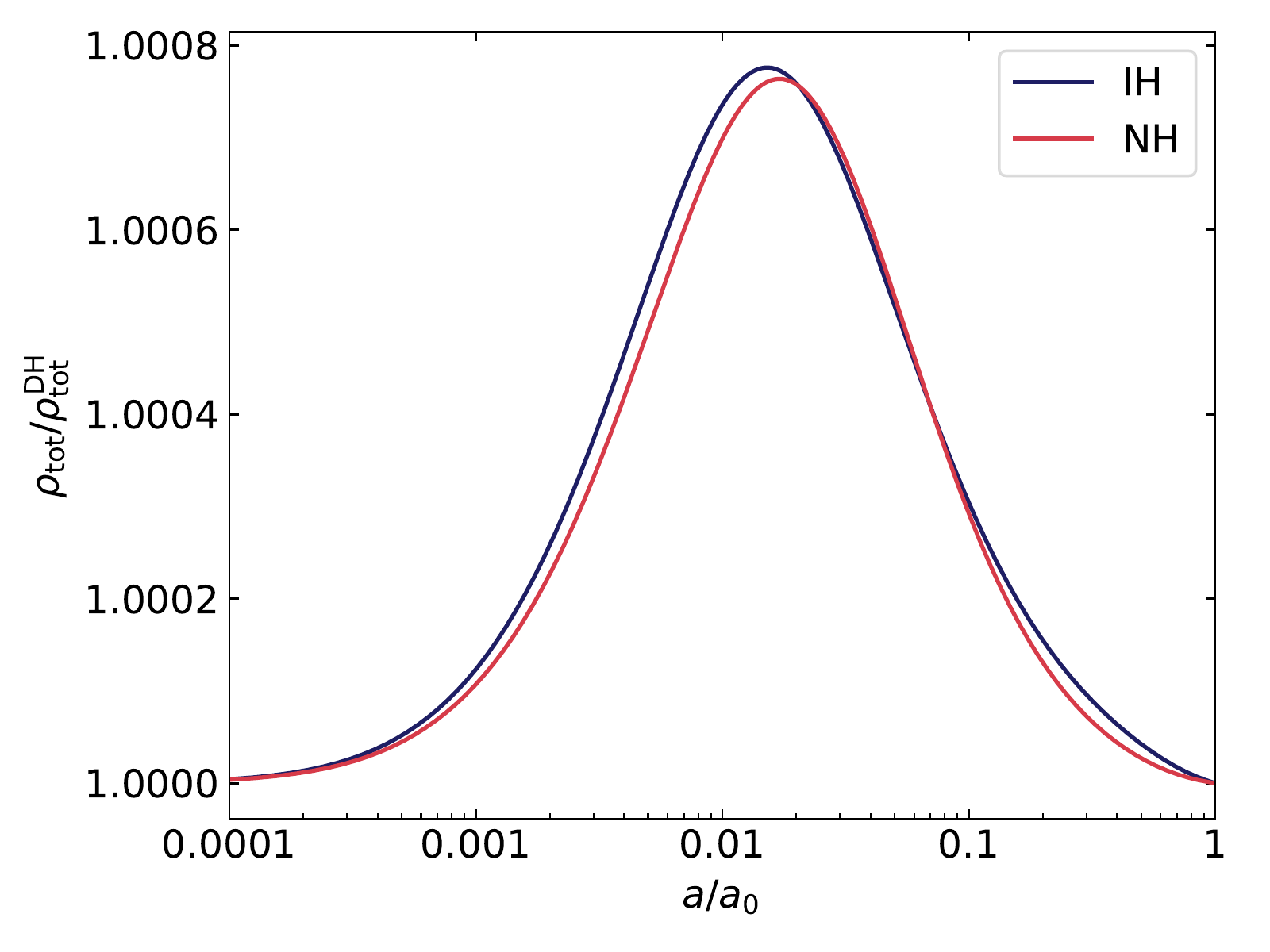}}\par
\caption{\label{fig:bkg} 
Fig~\ref{fig:bkg:nu}: Ratios of neutrino energy density and $\rho+3P$ for NH (red) and IH (blue) versus their respective DH, plotted as functions of scale factor.
Fig~\ref{fig:bkg:tot}: Ratio of total density for the same models.
}
\end{figure}

It is of interest to check the difference between hierarchical and degenerate neutrinos with {\it the same} contribution to the asymptotic late-time density, i.e. with $\sum m_\nu$ equal in the two cases.
In terms of the background expansion rate the quantities of interest are the neutrino energy density, $\rho$, which enters the Friedmann equation, and the quantity $\rho+3P$, which enters the acceleration equation.

In terms of the energy density the ratio is given by
\begin{equation}
\frac{\rho_{\rm NH,IH}}{\rho_{\rm DH}} = \frac{\sum_i \int d^3p \sqrt{p^2+m_i^2} f(p)}{3  \int d^3p \sqrt{p^2+m_D^2} f(p)},
\end{equation}
and for the quantity $\rho+3P$ we have
\begin{equation}
\frac{(\rho+3P)_{\rm NH,IH}}{(\rho+3P)_{\rm DH}} = \frac{\sum_i \int d^3p \left(\sqrt{p^2+m_i^2}+p^2/\sqrt{p^2+m_i^2}\right) f(p) }{3  \int d^3p \left(\sqrt{p^2+m_D^2}+p^2/\sqrt{p^2+m_D^2}\right) f(p)}.
\end{equation}

In Fig.~\ref{fig:bkg:nu} we plot\footnote{The plots of this section have been produced using CLASS~\cite{Blas:2011rf} and cross-checked using CAMB~\cite{Lewis:1999bs}.} these two ratios as a function of the scale factor, while Fig.~\ref{fig:bkg:tot} shows the ratio of the total density (summed over baryons, dark matter, neutrinos, photons and $\Lambda$) for the same models. In order to fix the timeline, we remind the reader that the redshift of the non-relativistic transition is given by
$a_\mathrm{nr}/a_0 \simeq 5.3 \times 10^{-4} \left( 1\,\mathrm{eV} / m_i \right)$, and that neutrinos with $m_i \lesssim 0.6$ eV go non-relativistic after recombination. As can be seen, it is always the case that energy density and pressure are higher for NH or IH than for DH during the transition from relativistic to non-relativistic, while in both the $T \to \infty$ and the $T \to 0$ limit they asymptote to the same values.
We would therefore expect the effect of the NH, IH versus DH treatment to primarily show up during this period. Since photon decoupling occurs at  $a/a_0 \sim 10^{-3}$, effects on the CMB primary signal should be minimal, while effects on structure formation should accumulate during the transition.

The NH model and its equivalent DH model start to differ when NH has one neutrino becoming non-relativistic while DH has none. Instead, the IH model and its DH equivalent start to differ when IH has {\it two} neutrinos becoming non-relativistic while DH has none. Thus, at early times, the IH/DH density ratio is larger than the NH/DH density ratio (both in terms of neutrino density only and in terms of total density). This appears as a small feature in Fig.~\ref{fig:bkg:nu}, but it is better seen in  Fig.~\ref{fig:bkg:tot} for $a/a_0 \leq 10^{-2}$. Indeed, when neutrinos are non-relativistic, they only enhance the total matter density by a very small amount, of the order of $\Omega_\nu^i$ (where $i \in 1,2,3$ is the index of mass eigenstates); but when they are still relativistic, like at the beginning of matter domination, they enhance it much more, by a factor of order $(a_\mathrm{nr}^i/a)\Omega_\nu^i$. Thus, the differences appearing between the various models at early times have more weight in the evolution of the total density than the differences appearing at late times. The fact that the IH/DH total density ratio is slightly larger than the NH/DH one for a long period of time will play a role in the discussion of the next sections.

\subsection{Free-streaming scales}

\begin{figure}[tbp]
\centering
\subfloat[]{\label{fig:fs_nh}\includegraphics[width=.5\linewidth]{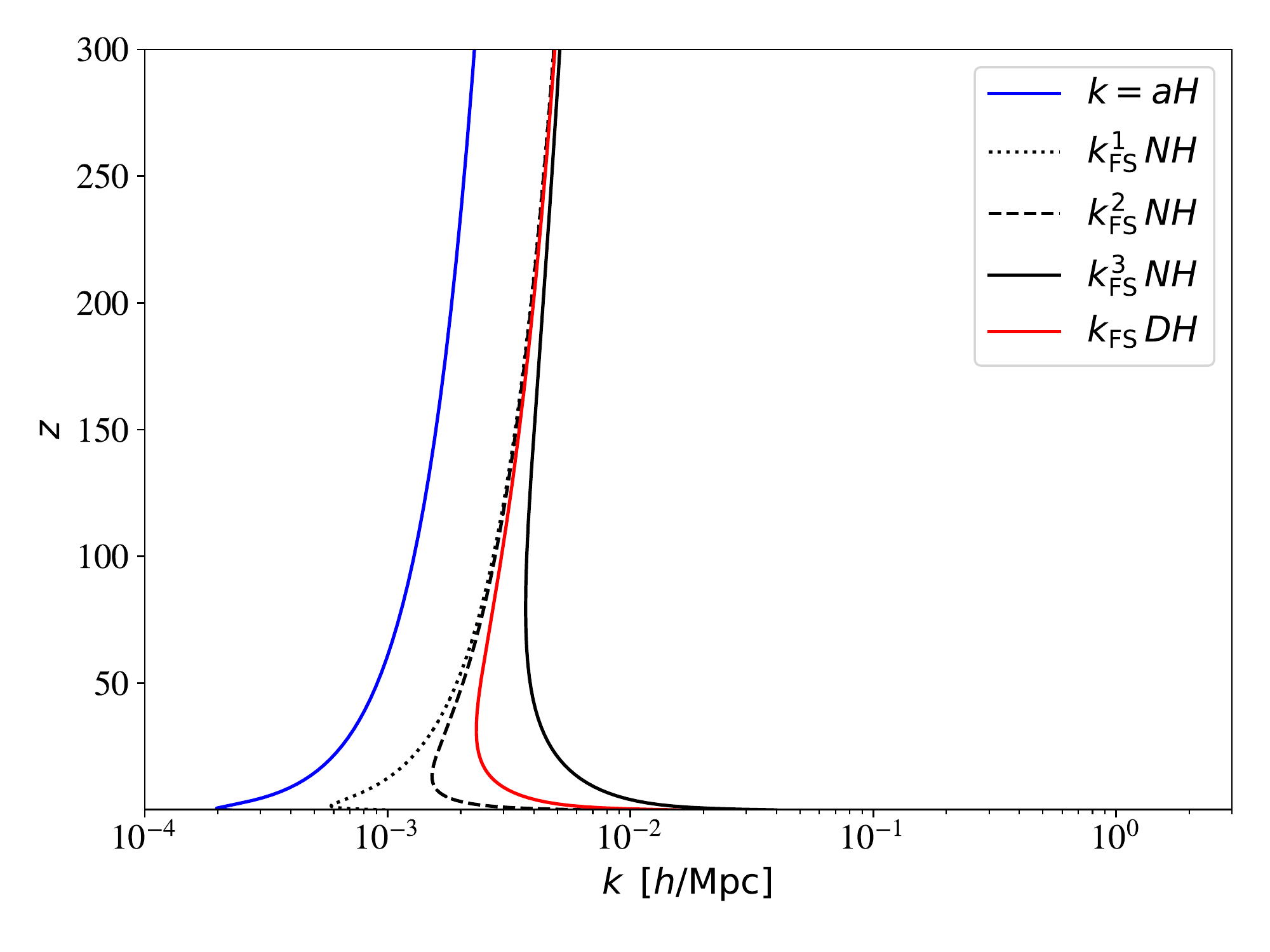}}\hfill
\subfloat[]{\label{fig:fs_ih}\includegraphics[width=.5\linewidth]{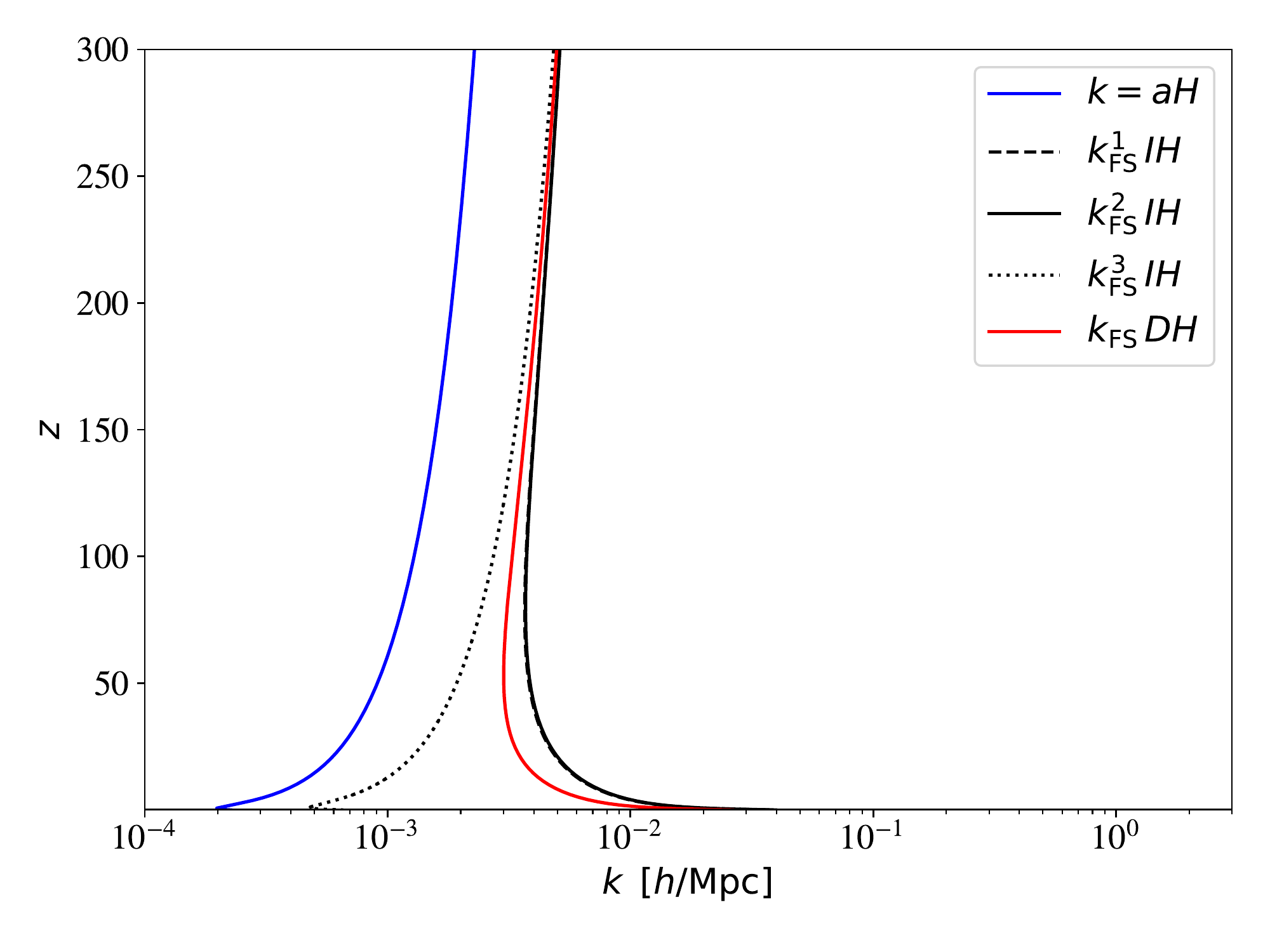}}\par
\caption{
Evolution of various characteristic wavenumbers at different redshifts. The scale $k=aH$ gives the order of magnitude of the Hubble crossing scale. The instantaneous free-streaming scales $k_\mathrm{FS}(z)$ are shown for each mass eigenstates, for the NH model and its DH equivalent (\ref{fig:fs_nh}), or for the IH model and its DH equivalent (\ref{fig:fs_ih}).
\label{fig:fs}
}
\end{figure}

There are several definitions of the neutrino-free streaming scales in the literature. 
\begin{itemize}
\item
The Free-Streaming Horizon (FSH) is the maximum distance over which free--streaming neutrinos can travel between their decoupling and today. It derives from an integral over the neutrino velocity (see e.g.\cite{Lesgourgues:1519137,Brandbyge:2017tdc}), and it can be associated to a wavenumber $k_\mathrm{FSH}$. Physically, it gives the largest wavelength at which perturbations are potentially affected by neutrino free-streaming effects.
\item
The instantaneous Free--Streaming (FS) length is a quantity similar to the Jeans length of a fluid, 
\begin{equation}
\lambda=2\pi\sqrt{\frac{2}{3}}\frac{c_s}{H}~,
\label{eq:fs}
\end{equation}  
that defines the region in which neutrino perturbations decay: thus it also gives an indication on where and when the growth rate of CDM perturbations is reduced. It can be associated to a function $k_\mathrm{FS}(z)$, that has a turnover at the time of the neutrino non-relativistic transition. The quantity $k_\mathrm{FS}(z=0)$ is much larger than $k_\mathrm{FSH}$ and is a good approximation for the scale at which neutrino free-streaming effects saturate in the linear matter power spectrum at redshift zero.
\item
Finally, the free-streaming scale evaluated at the redshift of the neutrino non-relativistic transition $z_\mathrm{NR}$ gives a proxy for the free-streaming horizon. It is associated to the minimum free-streaming wavenumber $k_\mathrm{NR}=k_\mathrm{FS}(z_\mathrm{NR})$.
\end{itemize}
In Fig.~\ref{fig:fs} we show the instantaneous free--streaming scale $k_\mathrm{FS}(z)$ (defined as in Eq.~(\ref{eq:fs}) where $c_s$ is the neutrino sound speed) for each neutrino species, for the four NH, IH and DH models considered in this discussion. In each model, the most important role is played by the smallest free-streaming length, i.e. the one associated to the heaviest eigenstate, shown on the plot with solid lines for each model.

\subsection{Changes to perturbations}

To understand the impact of the different scenarios, we can concentrate on the equation of evolution of CDM density fluctuations in the Newtonian gauge,
\begin{equation}
\delta_\mathrm{cdm} '' =
- k^2 \psi
- \frac{a'}{a} \delta_\mathrm{cdm} '
+ 3 \left( \phi'' + \frac{a'}{a} \phi'\right)~,
\label{eq:delta_cdm}
\end{equation}
(see e.g. \cite{Hannestad:2006zg,Lesgourgues:2006nd,Wong:2011ip,Lesgourgues:1519137}).
The three terms on the right-hand side account respectively for
\begin{itemize}
\item gravitational forces, which are responsible for gravitational clustering and for the growth of $\delta_\mathrm{cdm}$ on sub-Hubble scales; 
\item Hubble friction, which slows down gravitational clustering (because all distances between overdensities get stretched by the expansion); 
\item local density dilation effects (since $\phi$ represents a local modulation of the scale factor).
\end{itemize}

\begin{figure}[tbp]
\centering
\subfloat[]{\label{fig:k2_ratio}\includegraphics[width=.5\linewidth]{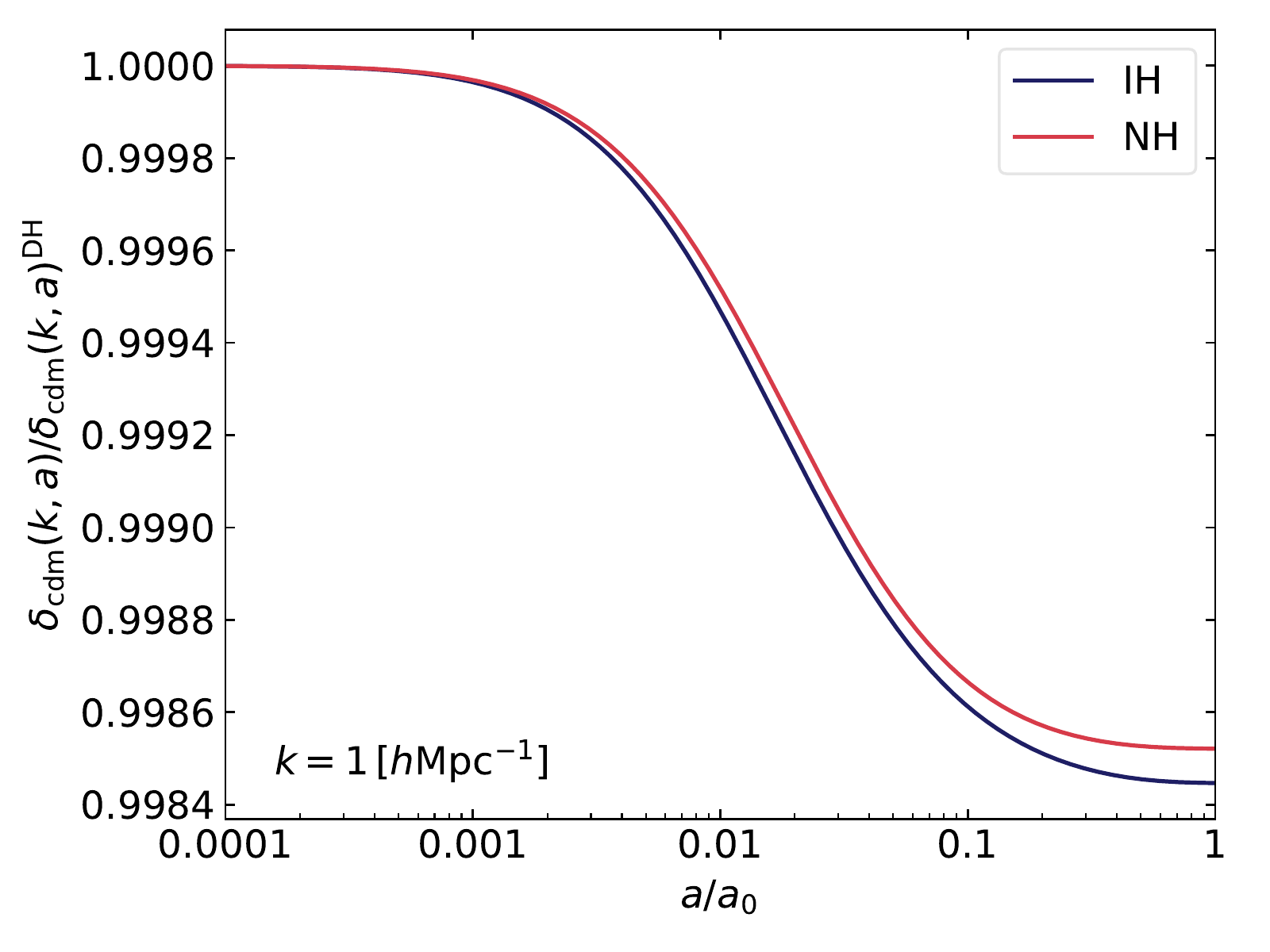}}\hfill
\subfloat[]{\label{fig:k0_ratio}\includegraphics[width=.5\linewidth]{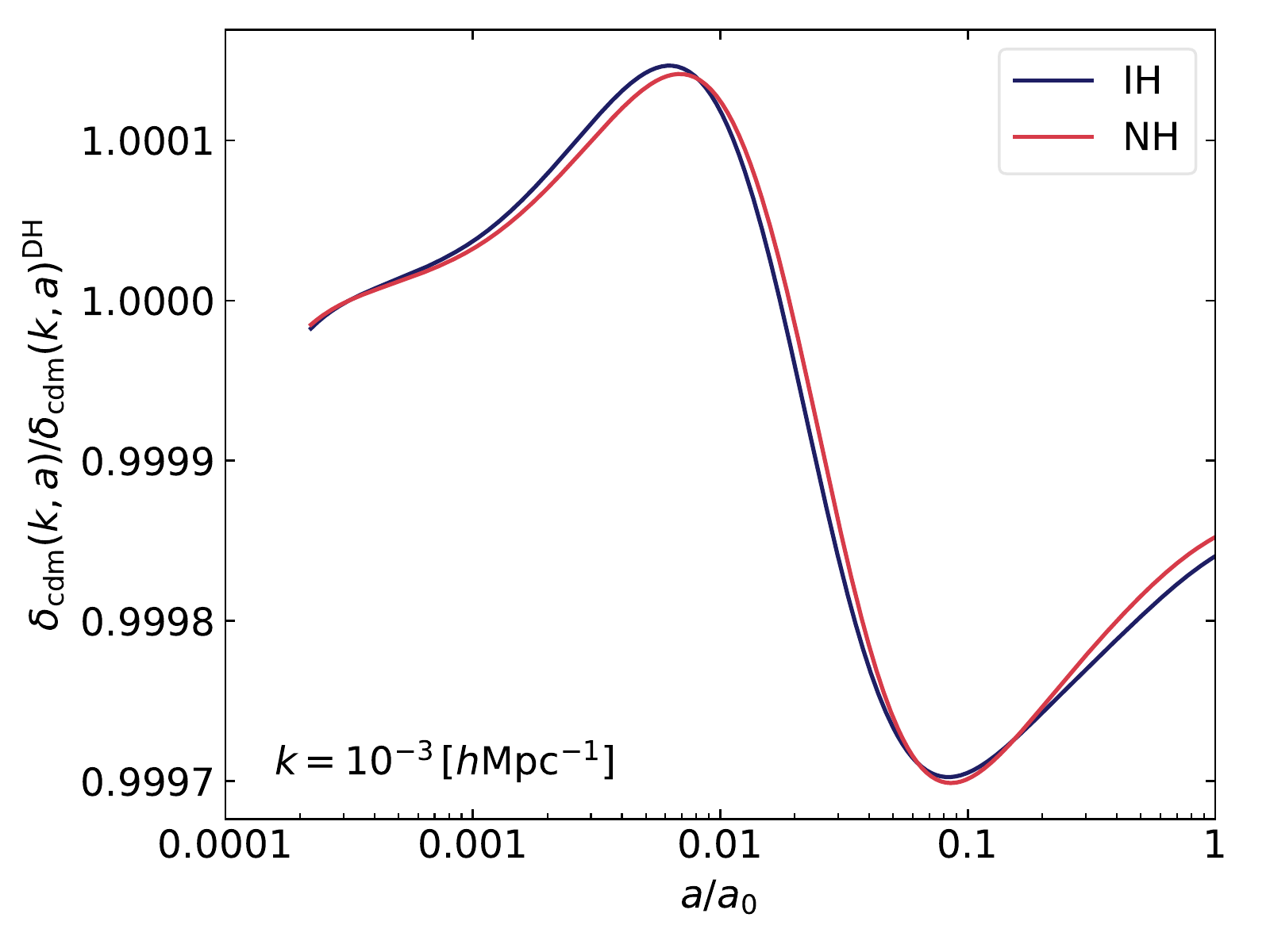}}\par 
\subfloat[]{\label{fig:tk_ratio}\includegraphics[width=.5\linewidth]{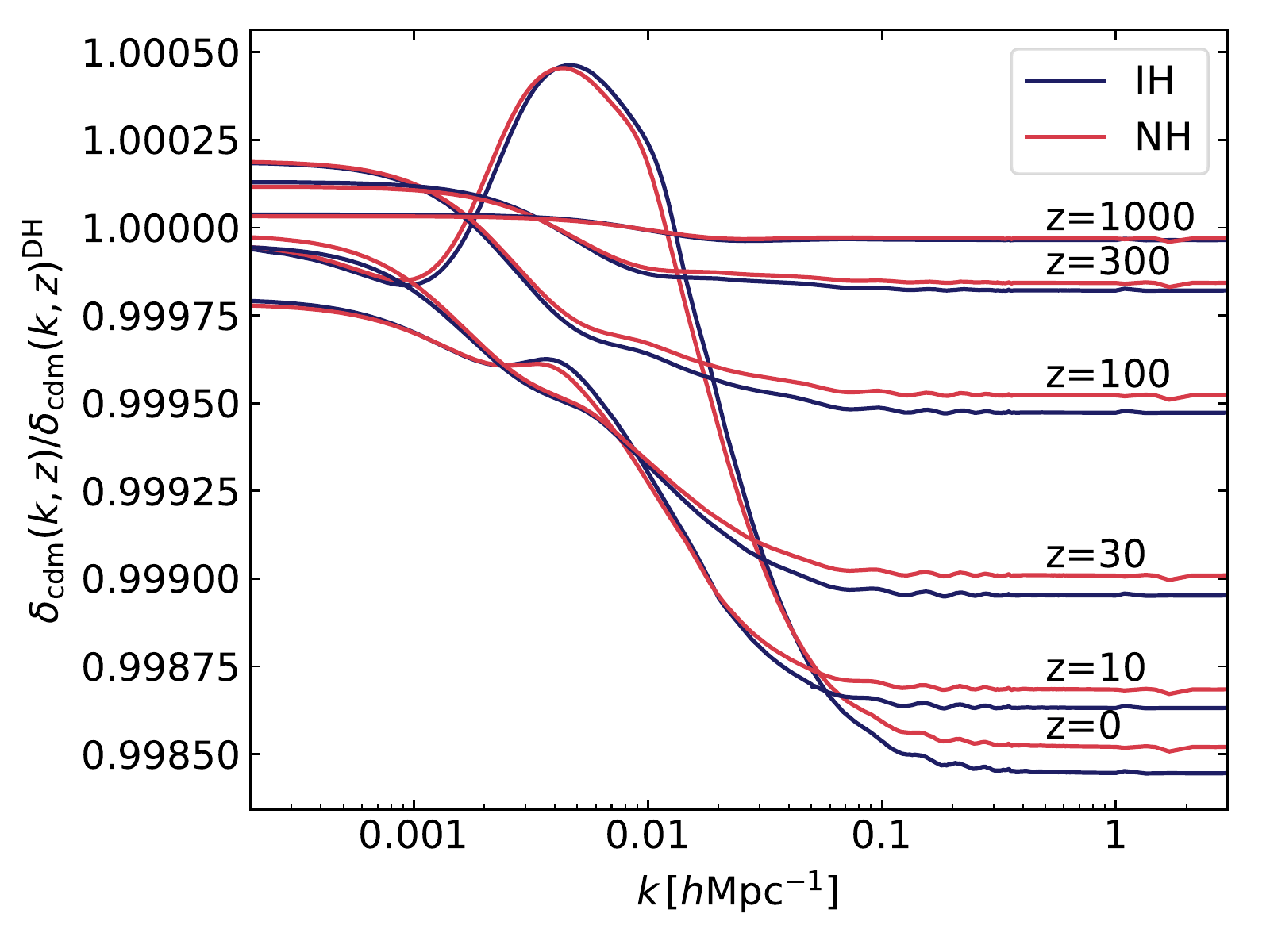}}\hfill
\subfloat[]{\label{fig:tk_ratio_nodilation}\includegraphics[width=.5\linewidth]{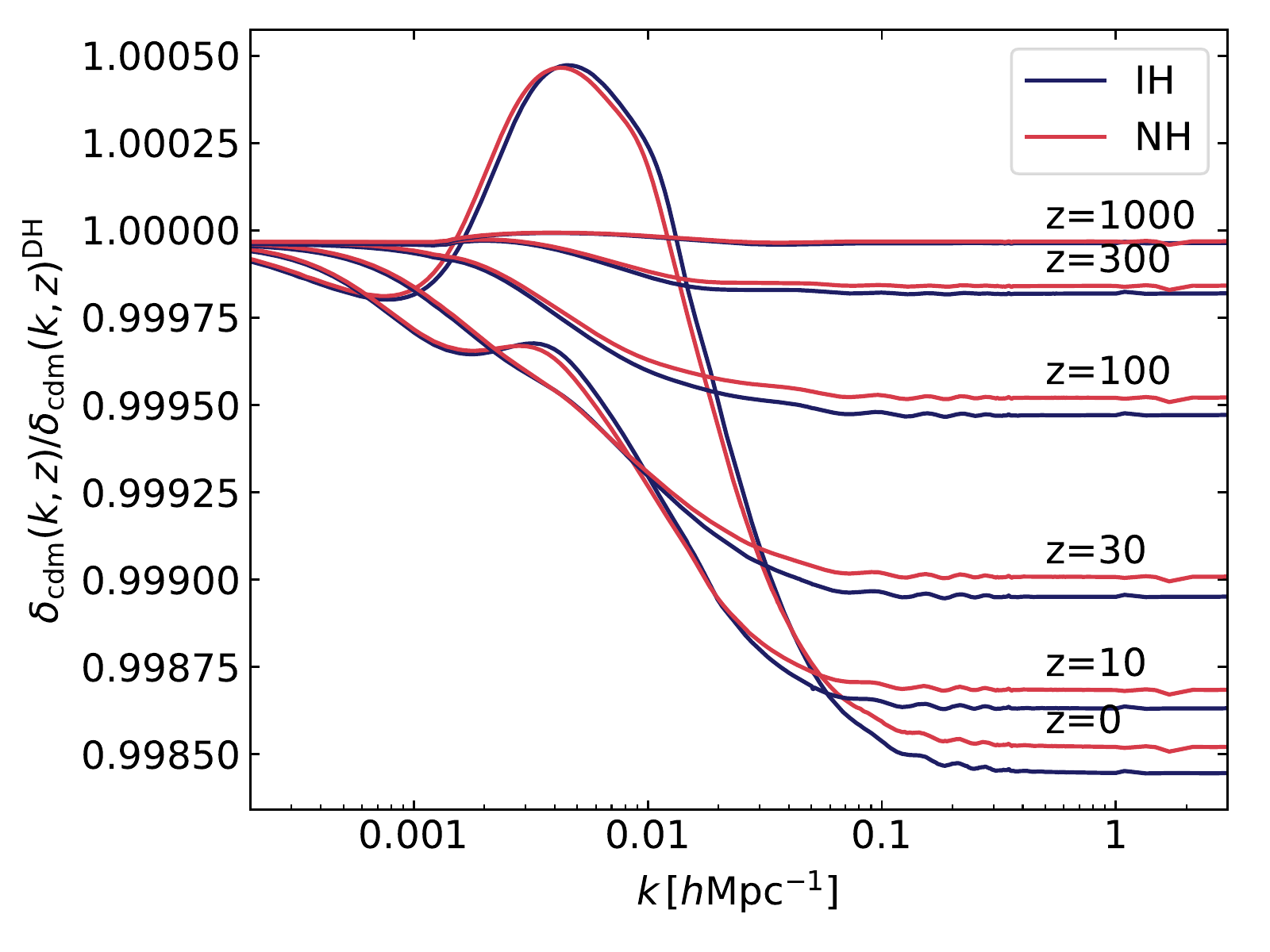}}\par 
\caption{
Fig.~\ref{fig:k2_ratio}: Ratio of CDM perturbation $\delta_\mathrm{cdm}(a,k)$ for NH (red) or IH (blue) compared to their  respective equivalent DH model, plotted as a function of $a$ for $k=1\,h\,\mathrm{Mpc}^{-1}$;
Fig.~\ref{fig:k0_ratio}: Same for $k=10^{-3}\,h\,\mathrm{Mpc}^{-1}$; 
Fig.~\ref{fig:tk_ratio}: Same ratio, plotted as a function of $k$ for different redshifts;
Fig.~\ref{fig:tk_ratio_nodilation}: Same when the dilation term is neglected in the equation of evolution of $\delta_\mathrm{cdm}$.
\label{fig:perturbations}
}
\end{figure}

Fig.~\ref{fig:k2_ratio} shows the evolution of CDM density perturbations in the NH or IH model compared to the equivalent DH model, for a small wavelength corresponding to $k=1\,h\,\mathrm{Mpc}^{-1}$. The ratio of perturbations drops below one when $a/a_0>10^{-3}$, i.e. after redshift $1000$. This can be easily understood analytically. At such small wavelengths, Hubble crossing takes place at a time when all neutrinos are ultra-relativistic and the NH/IH/DH models are all equivalent (the mode shown in Fig.~\ref{fig:k2_ratio} enters the Hubble scale at redshift 4800). Inside the Hubble radius, the dilation term of Eq.~(\ref{eq:delta_cdm}) becomes negligible, while neutrino perturbations are strongly suppressed compared to CDM perturbations. Then, using the Poisson limit of Einstein equations, we can simplify Eq.~(\ref{eq:delta_cdm}) into
\begin{equation}
\delta_\mathrm{cdm} '' =
- \frac{a'}{a} \delta_\mathrm{cdm} '
+ 4 \pi G a^2 \left( \bar{\rho}_\mathrm{cdm}+\bar{\rho}_\mathrm{b} \right) \delta_\mathrm{cdm}~.
\label{eq:delta_cdm2}
\end{equation}
Since the CDM and baryon background densities are identical in all models, the difference between NH, IH and DH only shows up at the level of the Hubble friction term. We have seen in Fig.~\ref{fig:bkg:tot} that in the redshift range $0<z<1000$, the total background density gets slightly enhanced in the NH model, and even more in the IH model. This applies also to the expansion rate and to the Hubble friction term, since $3 (a'/a)^2 = 8 \pi G a^2 \bar{\rho}_\mathrm{tot}$. Thus, for $0<z<1000$, the growth rate of $\delta_\mathrm{cdm}$ gets reduced in the NH case, and even more in the IH case. This is exactly what we observe in Fig.~\ref{fig:k2_ratio}.\\

Fig.~\ref{fig:k0_ratio} shows the evolution of a mode in the opposite limit of a very large wavelength, $k=10^{-3}h\,\mathrm{Mpc}^{-1}$, that is still slightly above the Hubble radius at $z=0$. In this case, an important role is played by the dilation term of Eq.~(\ref{eq:delta_cdm}). On super-Hubble scales, $\phi$ varies each time that the equation of state of the universe changes. Thus the dilation term of Eq.~(\ref{eq:delta_cdm}) triggers a small variation of all matter fluctuations (of CDM, baryons, neutrinos) each time that a neutrino species becomes non-relativistic. Since the NH, DH and IH models feature several such transitions at redshifts between $z\sim 100$ and $z\sim 0$, $\delta_\mathrm{cdm}$ has a small evolution even on super-Hubble scales during matter domination, that is different for each neutrino mass model.\\

To get a summary of all differences in the evolution of each mode, we show in Fig.~\ref{fig:tk_ratio} the ratio of the CDM perturbations $\delta_\mathrm{cdm}(z,k)$ as a function of $k$ at different redshifts $0<z<1000$. Since the effect of the dilation term on the behavior of large wavelengths is rather complicated, we also present in Fig.~\ref{fig:tk_ratio_nodilation} a version of the same plot obtained after modifying the equation of evolution of CDM perturbations in the Boltzmann code: we removed the dilation term from Eq.~(\ref{eq:delta_cdm}). In that case, the situation is clear:
\begin{itemize}
\item on super-Hubble scales, the perturbations are identical in the NH/IH/DH case;
\item on sub-Hubble scales, they are suppressed in the NH case, and even more in the IH case; since this suppression is caused by the excess of Hubble friction in the redshift range $0<z<1000$, it saturates for modes that crossed the Hubble radius before $z\sim 1000$, i.e. approximately for $k>0.04h\,\mathrm{Mpc}^{-1}$.
\item on top of this, a bump appears after $z\sim 30$ around the scale $k=4 \times 10^{-3}h\,\mathrm{Mpc}^{-1}$; it keeps increasing with time, such that at $z=0$ fluctuations on these scales are larger in the NH/IH cases than in the DH case.
This behavior affects the scales that lay between the shortest free-streaming length of the NH/IH model and that of the equivalent DH model, located precisely around $k=4 \times 10^{-3}h\,\mathrm{Mpc}^{-1}$ according to Fig.~\ref{fig:fs}. In this range, in the NH/IH case, the wavelength never enters into the neutrino free-streaming region, and CDM clusters as fast as in a massless model; while in the equivalent DH case, wavelengths do enter into this region, and the CDM growth rate is slightly reduced. At small redshift and near $k=10^{-2}h\,\mathrm{Mpc}^{-1}$, this effect wins over that of the Hubble friction, and the ratio of CDM perturbations gets inverted.
\end{itemize}
In Fig.~\ref{fig:tk_ratio}, the full equation has been used, including the dilation term. The picture is identical on all small (and in principle observable) wavelengths, while at very large scales, the behavior becomes more complicated due to dilation effects at each neutrino non--relativistic transition.\\

\begin{figure}[tbp]
\centering
\subfloat[]{\label{fig:pk_ratio_1}\includegraphics[width=.5\linewidth]{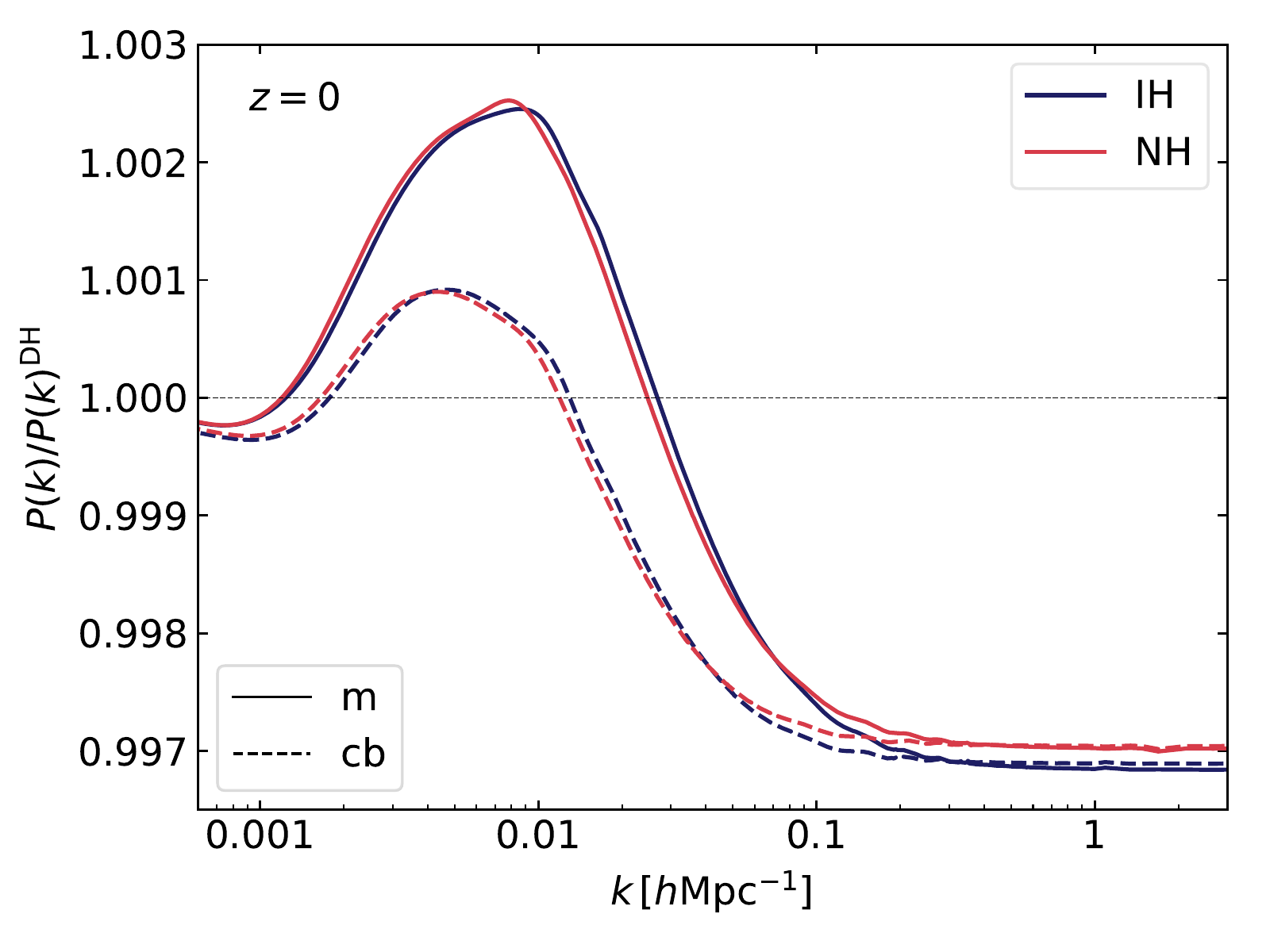}}\hfill
\subfloat[]{\label{fig:pk_ratio_2}\includegraphics[width=.5\linewidth]{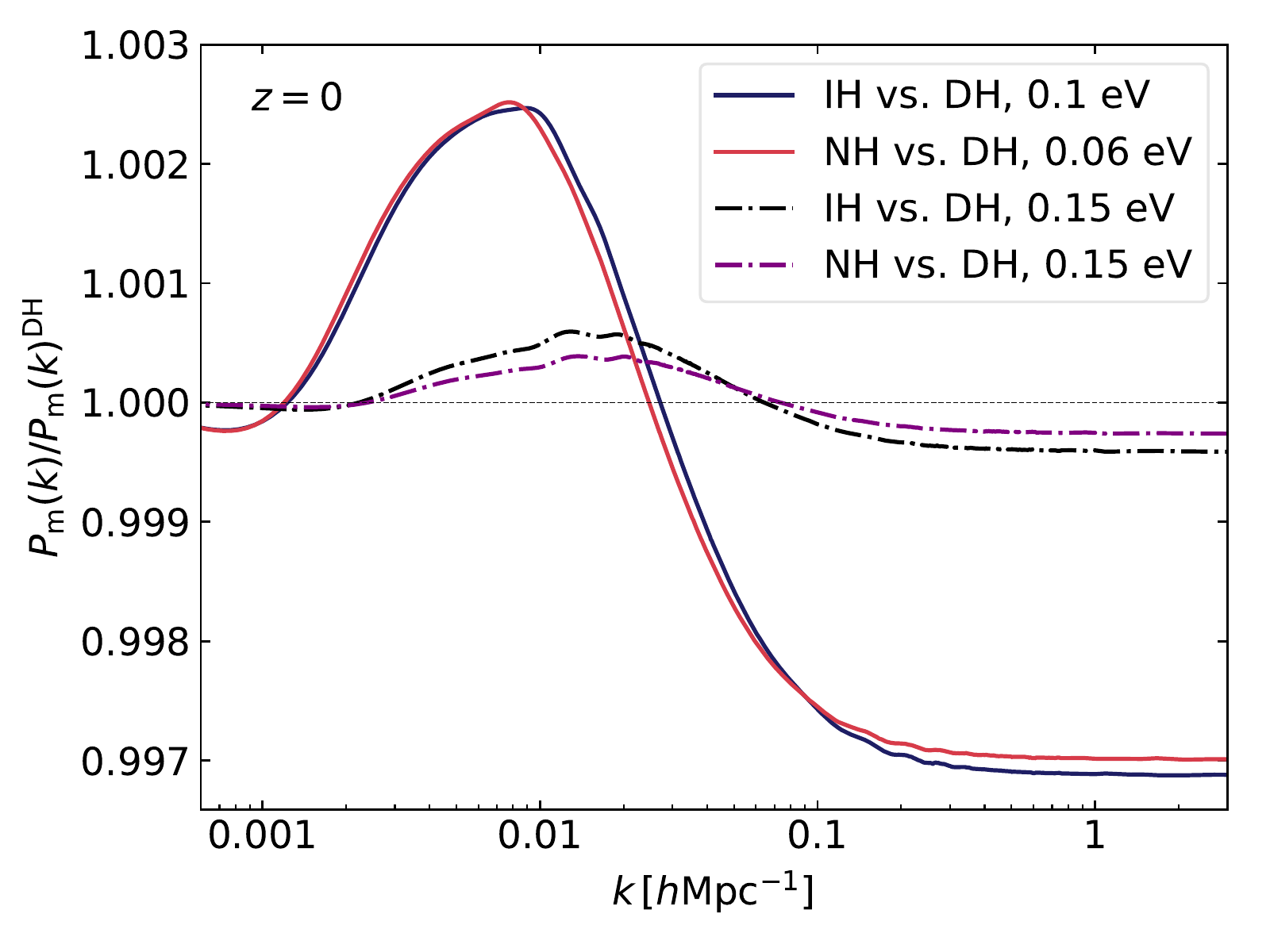}}\par 
\caption{
Fig.~\ref{fig:pk_ratio_1}: Ratio of total matter power spectrum $P_\mathrm{m}(k,z)$ (solid) and CDM plus baryon power spectrum $P_\mathrm{cb}(k,z)$ (dashed) for NH versus DH with $\sum m_\nu = 0.06$ eV (red) and IH versus DH with $\sum m_\nu = 0.10$ eV (blue) as a function of $k$ and at redshift $z=0$.
Fig.~\ref{fig:pk_ratio_2}: Ratio of total matter power spectrum $P_\mathrm{m}(k,z)$ for NH versus DH (purple) and IH versus DH (black) for a larger neutrino mass sum of $0.15$ eV (dot-dashed lines). For comparison we show again the $P_\mathrm{m}(k,z)$ ratio for the same minimal masses as in Fig.~\ref{fig:pk_ratio_1} (solid lines).
\label{fig:pk_ratio}
}
\end{figure}

In Fig.~\ref{fig:pk_ratio_1}, we show the ratio of the total matter power spectrum $P_\mathrm{m}(k,z=0)$ in the NH/IH versus DH model at redshift zero, as well as the ratio of the CDM plus baryon component only, $P_\mathrm{cb}(k,z=0)$. The latter behaves like the square of the CDM transfer function $\delta_\mathrm{cdm}(k,z=0)$, while the former has a larger bump on scales close to $k=10^{-2}h\,\mathrm{Mpc}^{-1}$. Indeed, since this scale is the one above which neutrino perturbations are not suppressed by free-streaming, it is also the one above which the total matter power spectrum receives contributions from both the neutrino and the CDM+baryon components. In the ratio of transfer functions, the neutrino component $\delta_\nu$ also has a bump, since on intermediate scales neutrinos are clustered in the NH/IH model and suppressed in the DH model. In the ratio of total matter power spectra, the bumps of the CDM+baryon and neutrino components add up and get squared, and the total ratio reaches about 0.25\%. On smaller scales, the suppression induced by the additional Hubble friction is around -0.30\%.

Finally, Fig.~\ref{fig:pk_ratio_2} shows the ratio of the total matter power spectrum $P_\mathrm{m}(k,z=0)$ in the NH/IH versus DH model both for the minimal mass allowed in each hierarchy (solid lines) and for a larger neutrino mass of $0.15$ eV (dot-dashed lines). When the neutrino mass sum increases, the amplitude of the bump and the relative suppression become smaller. This was expected since the small mass splittings of equations~(\ref{eq:split}) become negligible for a larger neutrino mass sum. Thus, for heavier neutrinos, the total energy density of the different cases deviate from each other by a smaller amount and during a shorter period of time than in Fig.~\ref{fig:bkg:tot}.

\subsection{Changes to CMB observables - with and without lensing}
\label{sec:cmb}
Given our previous considerations on background quantities and perturbations, we now discuss the impact on observables, starting from CMB.
\begin{figure}[h]
\centering 
\subfloat[]{\label{fig:cl}\includegraphics[width=.5\linewidth]{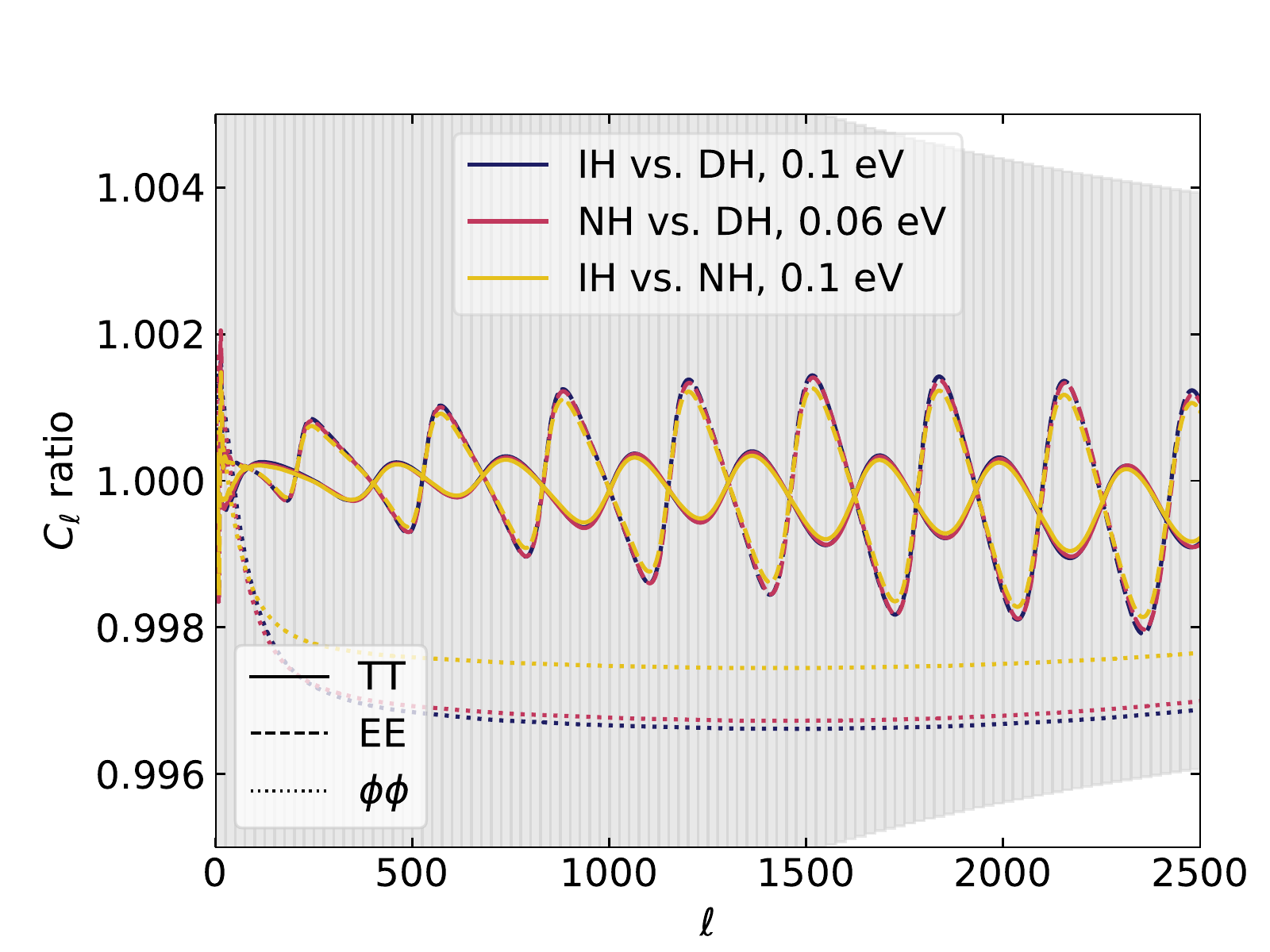}}\hfill
\subfloat[]{\label{fig:lensed_cl}\includegraphics[width=.5\linewidth]{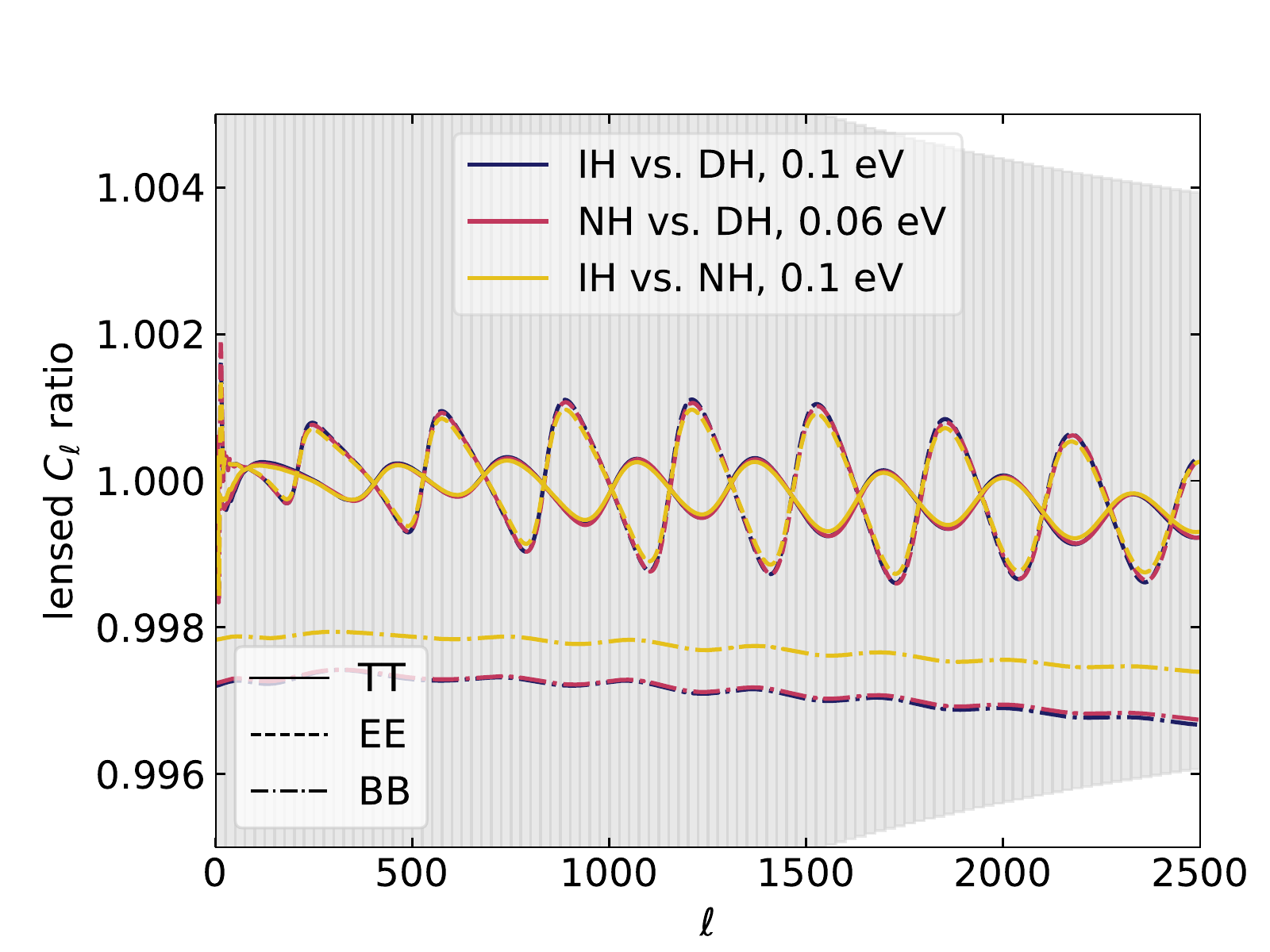}}\hfill
\label{fig:cmb}
\caption{Fig.~\ref{fig:cl}: Ratios of unlensed $C_\ell$ between NH and DH (red, $\sum m_\nu=0.06$ eV), between IH and DH (blue, $\sum m_\nu=0.10$ eV), and between IH and NH (yellow, $\sum m_\nu=0.10$ eV). The grey shaded area represents binned (100 bins in the range $2<\ell<2500$) cosmic variance. 
Fig.~\ref{fig:lensed_cl}: Same as Fig.~\ref{fig:cl}, but now including lensing.
}
\end{figure}

In Fig.~\ref{fig:cl} we show the ratio of unlensed  angular power spectra of temperature (T), E-mode polarization (E), and lensing potential ($\phi$) between the hierarchies (NH and IH) and the degenerate case (DH) with the same total mass ($0.06$ eV for NH and $0.10$ eV for IH).
First of all notice that the variations are well within cosmic variance, and, thus, unobservable.
The oscillatory behavior is due to a shift in the position of the CMB peaks, mainly caused by the variation of the angular diameter distance that follows the change in the background expansion at intermediate redshifts (see Fig.~\ref{fig:bkg}).
The amplitude of the oscillations is larger for polarization than for temperature, because the polarization spectrum only comes from acoustic oscillations on the last scattering surface, and receives no corrections from, e.g., the Doppler effect.
Since the lensing potential spectrum follows the trend of the matter power spectrum up to a smoothing kernel from Fourier to multipole space, the spectrum $C_l^{\phi \phi}$ is suppressed on small scales for IH and NH compared to the equivalent DH models: thus there is slightly less CMB lensing in the hierarchical cases.

In Fig.~\ref{fig:lensed_cl} we show the ratio of lensed angular power spectra of temperature (T), E-mode polarization (E) and B-mode polarization (B) for the same models. Since in the first place, the peaks are smoothed in the lensed spectra compared to the unlensed ones, the shift induced by the different neutrino splittings produces oscillations with a smaller amplitude.
For the B-mode spectra, the small oscillations are combined with an overall reduction of power in the NH/IH case. Indeed, in the absence of primordial tensor modes, the BB spectrum comes from a leakage from E-modes to B-modes induced by CMB lensing. The NH/IH models feature less CMB lensing and thus less leakage into the BB spectrum.

Finally, the yellow lines in Fig.~\ref{fig:cl} and in Fig.~\ref{fig:lensed_cl} show the ratio between the two hierarchies with the same total mass ($\sum m_\nu=0.10$ eV).
Notice that this ratio is smaller that the one between NH and DH (or IH and DH), with a reduction of the deviation visible mainly in the power spectra of lensing potential and of B-mode polarization. We will go back to this point in the next section.


\subsection{Changes to LSS observables - galaxy clustering and weak lensing}
\label{sec:lss}
We now turn our attention to the impact of neutrino mass ordering on LSS observables, which we expect to be more pronounced than on CMB spectra.
\begin{figure}[h]
\centering 
\subfloat[]{\label{fig:pkg}\includegraphics[width=.5\linewidth]{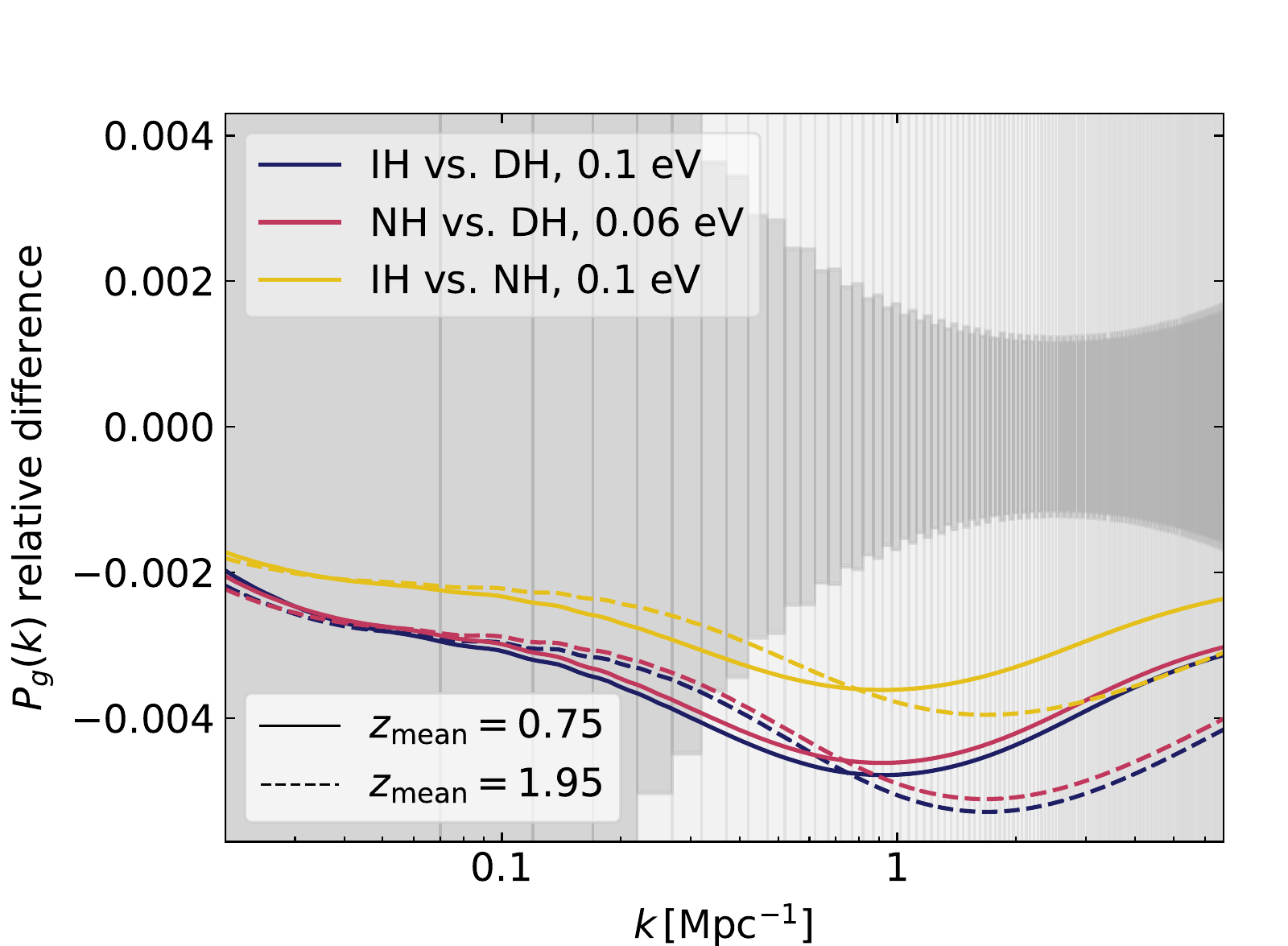}}\hfill
\subfloat[]{\label{fig:cls}\includegraphics[width=.5\linewidth]{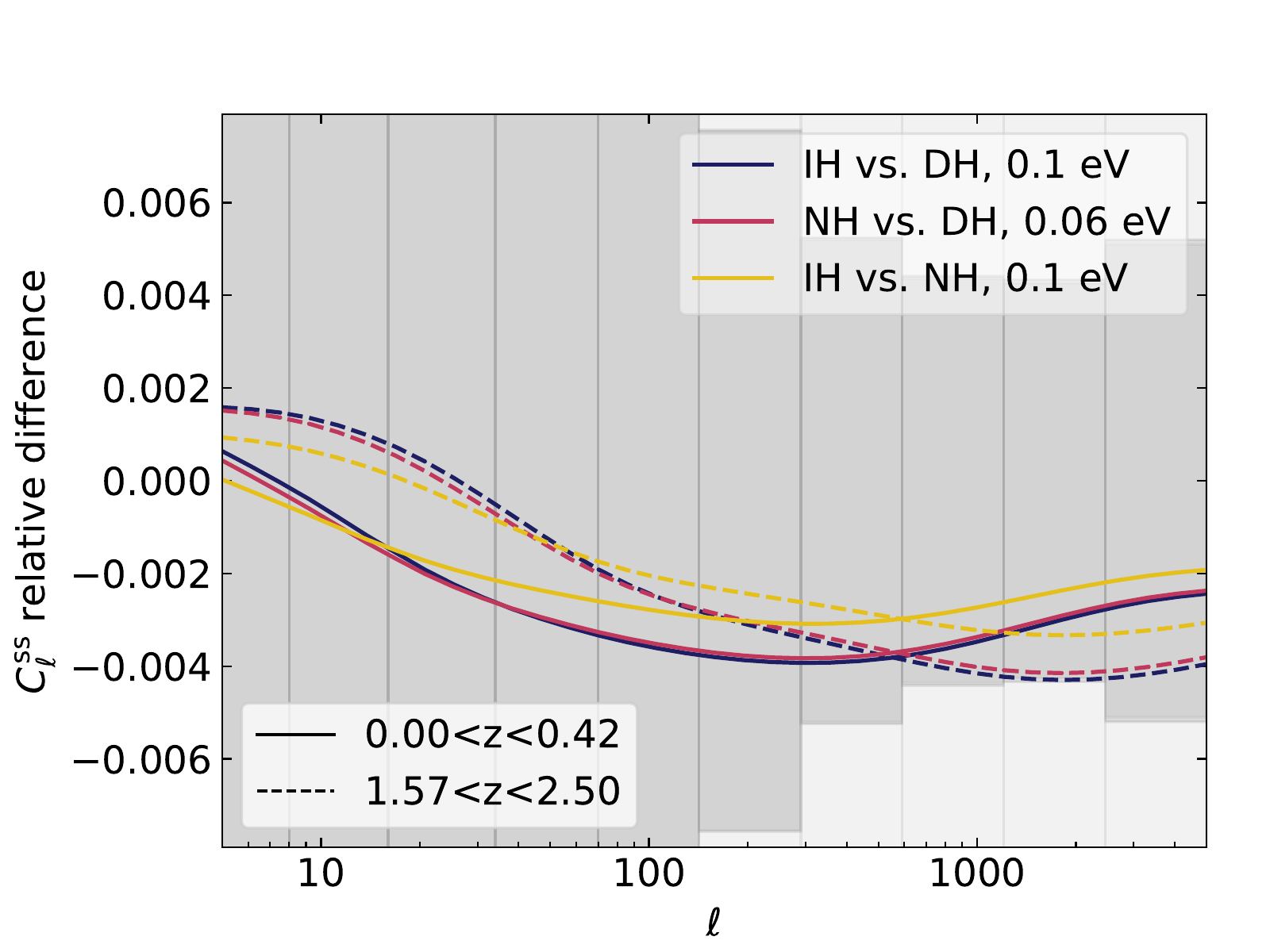}}\hfill
\label{fig:lss}
\caption{Fig.~\ref{fig:pkg}: Relative difference in the galaxy power spectrum along the line of sight ($\mu=0$) for NH versus DH with $\sum m_\nu = 0.06$ eV (red), IH versus DH with $\sum m_\nu = 0.10$ eV  (blue), and IH versus NH with $\sum m_\nu = 0.10$ eV (yellow), and for redshift bins centered on $z_\mathrm{mean}=0.75$ (solid lines) and $z_\mathrm{mean}=1.95$ (dashed lines).
We also show the effective observational error in bins of $\Delta k = 0.05\,\mathrm{Mpc}^{-1}$ for a Euclid-like survey (light grey shade for $z_\mathrm{mean}=1.95$ and dim grey shade for $z_\mathrm{mean}=0.75$), as defined in the text.
Fig.~\ref{fig:cls}: Relative difference in the cosmic shear angular power spectrum for the first and last redshift bin, assuming flat-sky and Limber approximations. The correspondence between models and colors is the same as in Fig.~\ref{fig:pkg}. The binned (with 11 equally-spaced bins in logarithmic scale) observational error for the last redshift bin is depicted in dim grey, while the one for the first redshift bin spans the whole y-axis (light grey).}
\end{figure}

The two main LSS observables of a future photometric and spectroscopic survey like Euclid \cite{Amendola:2016saw,Laureijs:2011gra} are galaxy clustering and weak lensing.
Galaxy clustering is modelled as a galaxy power spectrum $P_g(k,\mu,z)$, where $\mu$ is the angle between the line of sight and the direction in which the power spectrum is measured.
The galaxy power spectrum is proportional to the CDM and baryon only power spectrum $P_{cb}(k,z)$ up to various correction factors (scale-independent linear bias \cite{Castorina:2013wga,Raccanelli:2017kht,Vagnozzi:2018pwo}, redshift-space distortions, resolution effects, etc.).
Weak lensing is instead modelled as an angular power spectrum $C_\ell^{ij}$ with a source function given by the convolution of the matter power spectrum $P_m(k,z)$ with a window function depending on the galaxy distribution in the $i$ redshift bin.

Figs.~\ref{fig:pkg} and \ref{fig:cls} show the deviations of the hierarchies (NH - red and IH - blue) with respect to their respective reference DH cases in terms of $P_g(k,\mu,z)$ and $C_\ell^{ij}$.
We see less power in the NH and IH cases, which reflects the relative suppression of the 
total and CDM+baryon power spectra seen in Fig.~\ref{fig:pk_ratio_1}.
The shape of the step-like suppression is more evident in Fig.~\ref{fig:pkg} than in Fig.~\ref{fig:cls} because the convolution of the matter power spectrum with the window function redistributes power at different scales. For this reason, there is even a small relative increase in power in the shear angular power spectrum at very low $\ell$: this is connected to the bump in the 
matter power spectrum  (Fig.~\ref{fig:pk_ratio_1}) caused by the different free--streaming scales in the various models.
Besides the step-like suppression, there is a further dip at scales $\sim 1 {\rm Mpc}^{-1}$. This additional suppression is due to non-linear effects embedded in $P_{cb}(k,z)$ \cite{Takahashi:2012em} for $P_g(k,\mu,z)$, and in $P_m(k,z)$ \cite{Bird:2011rb} for $C_\ell^{ij}$. 
Non-linear effects are indeed sensitive to the change in the background caused by the different distribution of the total neutrino mass among the mass eigenstates.

The grey shades of Fig.~\ref{fig:pkg} and of Fig.~\ref{fig:cls} depict the effective observational errors expected for a future Euclid-like survey\footnote{By definition, this effective observational error is normalised in such way to have a very visual and intuitive meaning: a residual going through each error edge in one single bin (among all $k$-bins, $\mu$-bins, $z$-bins) would correspond to a model raising the total effective $\chi^2$ of the experimental likelihood by one unit compared to the fiducial model. This effective error reads:
\begin{equation}
\sigma_{\text{eff}}(k,\mu,\bar{z}) = \left[ P_g(k,\mu,\bar{z}) + P_N(k,\mu,z) \right] \, \left[k^3\frac{V}{2(2\pi)^2} \log\left(\frac{k_{\rm max}}{k_{\rm min}}\right) *2 *N \right]^{-1/2}
\end{equation}
where $P_N$ is the noise spectrum, $V$ the survey volume, $[k_{\rm min}, k_{\rm max}]$ the reconstructed wavenumber range, and $N$ the number of redshift bins (see e.g. equations (2.1), (2.2.) in \cite{Audren:2012vy}).
}.
The relative difference between NH/IH and DH in terms of weak lensing angular power spectrum is well below the observational error.
On the contrary, in the galaxy power spectrum, the relative difference between NH/IH and DH, although very small ($\sim 0.4$\%), exceeds the observational error for the lowest redshift bin. Thus, one could naively think that it might be detectable.
However, the signal is located at scales where theoretical uncertainties dominate over the observational error. Indeed, at such small scales the clustering is non-linear and baryonic effects start to take over ($\sim 10$\% at $k\sim0.7\,{\rm Mpc}^{-1}$). Poor knowledge of these effects (among others) makes the modeling of $P_g$ with a $\sim 0.1$\% accuracy impossible\footnote{1\% is already a challenge, see Ref. \cite{Schneider:2015yka}}, and so is the detection of such tiny neutrino mass hierarchy effect.

Fig.~\ref{fig:pkg} and Fig.~\ref{fig:cls} also show the ratio of the observables between the IH and the NH cases with the same total mass $\sum m_\nu = 0.10$ eV. As we already noticed in the previous section about the CMB spectra, the difference between these cases is even smaller than the one between NH and DH (or IH and DH) with the same total mass set to the minimum of each hierarchy. Thus, if cosmology is not sensitive to the difference between NH (or IH) and DH, it will not be able to tell the difference between NH and IH.

Finally, here we do not discuss galaxy bias, whose scale dependence can be considered as a signature of the individual neutrino masses.
Anyhow, Refs.\ \cite{Fidler:2018dcy, Munoz:2018ajr} have recently demonstrated that even neutrino bias is not sensitive to the specific mass hierarchy (see however Ref.\ \cite{LoVerde:2014pxa}).

\section{Cosmological parameter estimation and evidence}
\label{sec:results}
The aim of our analysis is to investigate if (and how) cosmological surveys in the near or distant future will ever be able to detect the neutrino mass hierarchy.
More specifically, in the remainder of the paper we will try to answer the following question: 
Will cosmology ever be sensitive to the difference in the signature of NH/IH cases with respect to the DH case that we have just discussed?

To address this question, we produce synthetic datasets according to a fiducial cosmology with either NH or IH (fiducial parameter values are given in Tab.~\ref{tab:fiducialmcmc}). We then fit the mock data with a Bayesian parameter inference algorithm, assuming the same cosmology, but either with the correct neutrino mass hierarchy (NH or IH) or with neutrinos degenerate in mass (DH). This strategy was already used in \cite{Hamann:2012fe} and in section 7.1 of \cite{DiValentino:2016foa} to evaluate the parameter reconstruction bias induced by wrong neutrino mass splitting assumptions, but here we will also compute Bayesian evidence ratios to show whether it is possible to discriminate between the correct hierarchy and the DH approximation.
\begin{table*}[!h]
\begin{center}
\begin{tabular}{cccccccc} 
 \hline 
 $\omega_{b}$ &$\omega_{cdm}$ &$100\theta_s$ &$\ln (10^{10}A_{\rm s})$ &$ n_{\rm s}$ &$z_\mathrm{reio}$ & $\sum m_\nu$(NH) eV & $\sum m_\nu$(IH) eV\\
\hline 
 $0.02218$ &$ 0.1205$ &$1.04146$  &$3.0560$ &$0.9619$ &$8.24$ &$0.10$ ($0.06$) &$0.15$ ($0.10$) \\
\hline
\end{tabular}
\caption{Fiducial values of the free cosmological parameters.
}
\label{tab:fiducialmcmc}
\end{center}
\end{table*}
Notice that, unlike in Section~\ref{sec:formalism}, we choose some values of the fiducial total mass slightly larger than the minimum mass of each hierarchy. 
This choice allows to recover two-sided posteriors on the total mass not just for DH, but also for NH and IH.

We devise three cases, always combining future CMB data (CMB) with future weak lensing photometric observations (WL), future spectroscopic galaxy clustering surveys (GC), and neutral hydrogen 21-cm intensity mapping (IM).
\begin{description}
\item[Realistic:] This case roughly corresponds to the case CMB-S4+LiteBIRD+Euclid+SKA1-IM-B2 of Ref.\ \cite{Brinckmann:2018owf}. 
Concerning CMB-S4 \cite{Abitbol:2017nao,Abazajian:2016yjj} and LiteBIRD \cite{Matsumura:2013aja,Suzuki:2018cuy} we use the same assumptions of Ref.\ \cite{Abazajian:2016yjj}, i.e., LiteBIRD for $\ell \leq 50$, CMB-S4 for $\ell > 50$ over 40\% of the sky, and LiteBIRD in the remaining 30\% of the sky.
For the Euclid-like survey, weak lensing and galaxy clustering, as well as for SKA phase 1 intensity mapping band 2, we use the same specifications of the realistic case of Ref.\ \cite{Sprenger:2018tdb}. Concerning galaxy clustering and intensity mapping we model the 3D signal including Alcock-Paczynski effect, redshift space distortions and fingers of God. We include non linear corrections from Halofit \cite{Takahashi:2012em} and a theoretical error parameterizing our uncertainty on it, as well as on baryonic feedback and scale dependent bias. For Euclid-like galaxy clustering we use 13 redshift bins spanning the range $0.7<z<2.0$, while SKA phase 1, band 2 (operating in single dish mode) intensity mapping allows to cover the low redshift range $0.05<z<0.45$.
For Euclid-like cosmic shear the signal is projected into a 2D angular power spectrum, assuming flat-sky and Limber approximations. The redshift range up to $2.5$ is divided into $10$ equally populated redshift bins. As for galaxy clustering, we include non linear corrections from Halofit \cite{Bird:2011rb}. Here the theoretical uncertainty on small scales is accounted for using a redshift dependent cut-off in Fourier space, that is then converted into a bin dependent cut-off in multipole space.
Notice that this case, although labeled as ``realistic'', already contains some optimistic assumptions, such as the perfect foreground cleaning in the CMB future surveys, as well as the modeling of the bias tuned by only two additional nuisance parameters. However, here we are not aiming at producing realistic forecast of the sensitivity of future cosmological surveys (see Ref.\ \cite{Blanchard:2019oqi} for the official Euclid Fisher forecast).
\item[Optimistic:] With respect to the realistic case, here we assume that the observations are boosted by an improved experimental setup, and the systematic errors are under control. The changes are:
\begin{itemize}
\item[CMB:] CMB-S4+LiteBIRD are replaced by PICO~\footnote{\url{https://zzz.physics.umn.edu/ipsig/_media/pico_science_aas_v11.pdf}}, which is cosmic variance limited up to $\ell=3500$ in TT and up to $\ell=2500$ in EE
\item[WL:] the number of galaxies per square arc min is increased from 30 to 40
\item[GC:] the limiting flux is set to $0.5\times10^{-16} \mathrm{erg\,cm}^{-2}\mathrm{\,s}^{-1}$, 6 times smaller than the expected one from NISP (table 3 Model 1 of Ref.\ \cite{Pozzetti:2016cch}); the additional nuisance parameters accounting for uncertainties on bias modeling are removed
\item[IM:] the additional nuisance parameters accounting for uncertainties on bias modeling and on redshift dependence of hydrogen distribution are removed
\end{itemize}
\item[Extreme:] With respect to the optimistic case, here we assume that in WL, GC, and IM, there is no theoretical uncertainty on the modeling of non-linear clustering on small scales.
\end{description}

\begin{figure}[tbp]
\centering
\begin{tabular}{cc}
\includegraphics[width=.5\linewidth]{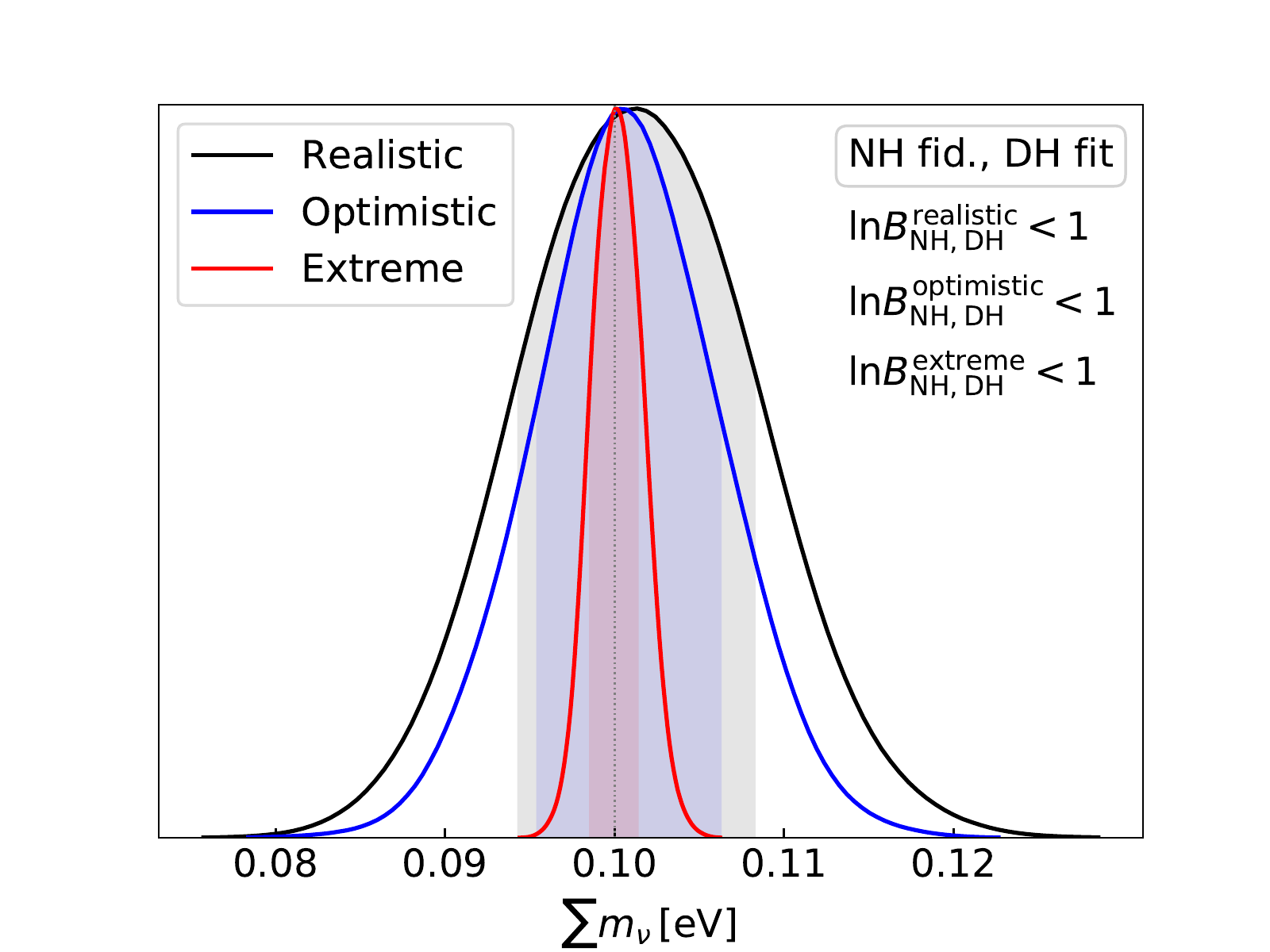}&\includegraphics[width=.5\linewidth]{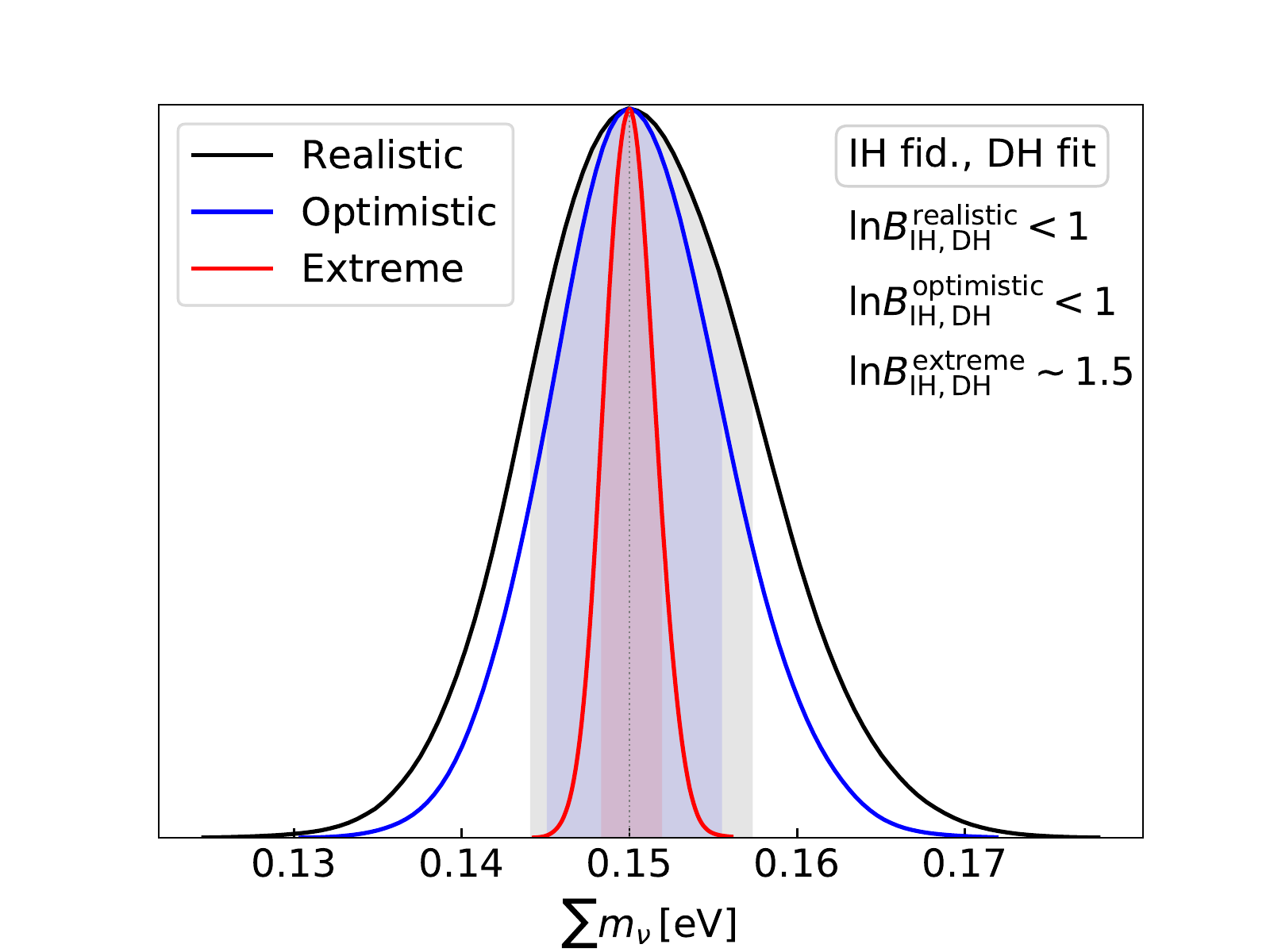}\\
\hfill
\end{tabular} 
\caption{\label{fig:posterior}Marginalized posterior distribution of the sum of neutrino masses for different combinations of datasets and specifications (see text for details): realistic (black), optimistic (blue), extreme (red). The fiducials are created assuming $\sum m_\nu=0.10$ eV and NH ({\it left panel}), and $\sum m_\nu=0.15$ eV and IH ({\it right panel}). The analysis is performed assuming the degenerate approximation $m_1=m_2=m_3$. Black dotted lines mark the fiducial $\sum m_\nu$. We also report the Bayes factors (ratios of the evidences). The shades represent the $1\,\sigma$ intervals.}
\end{figure}

For each of the ``realistic'', ``optimistic'' and ``extreme'' cases, we fitted NH to NH, DH to NH, IH to IH, and DH to IH. For each of these twelve fits, we produced MCMC chains with the package MontePython 3.1 \cite{Audren:2012wb,Brinckmann:2018cvx}. Additionally we used the MCEvidence package \cite{Heavens:2017afc} to compute the Bayes factors. We assumed flat priors on the parameters listed in table \ref{tab:fiducialmcmc}. Our prior boundaries are wide enough to ensure that the likelihood and each posterior probability become negligible on the two edges; this means that our results (including the calculation of the Bayes factors) do not depend explicitly on the prior edges.

As can be seen from Fig.~\ref{fig:posterior}, in all cases the Bayesian evidence factor of the model with the correct hierarchy is very close to that with the degenerate mass approximation, such that $|\ln (B_\mathrm{NH, IH}/B_\mathrm{DH})|<1$. This corresponds to  {\it inconclusive} on Jeffrey's scale \cite{Jeffreys}. Thus, for a fiducial total mass of $0.10$ or $0.15$ eV, the experimental data will never allow us to discriminate between the different hierarchies.
We find that the bias on the reconstruction of the neutrino mass sum induced by the approximate DH model is negligible, even in the ``extreme'' case.
\begin{figure}[tbp]
\centering
\begin{tabular}{cc}
\includegraphics[width=.5\linewidth]{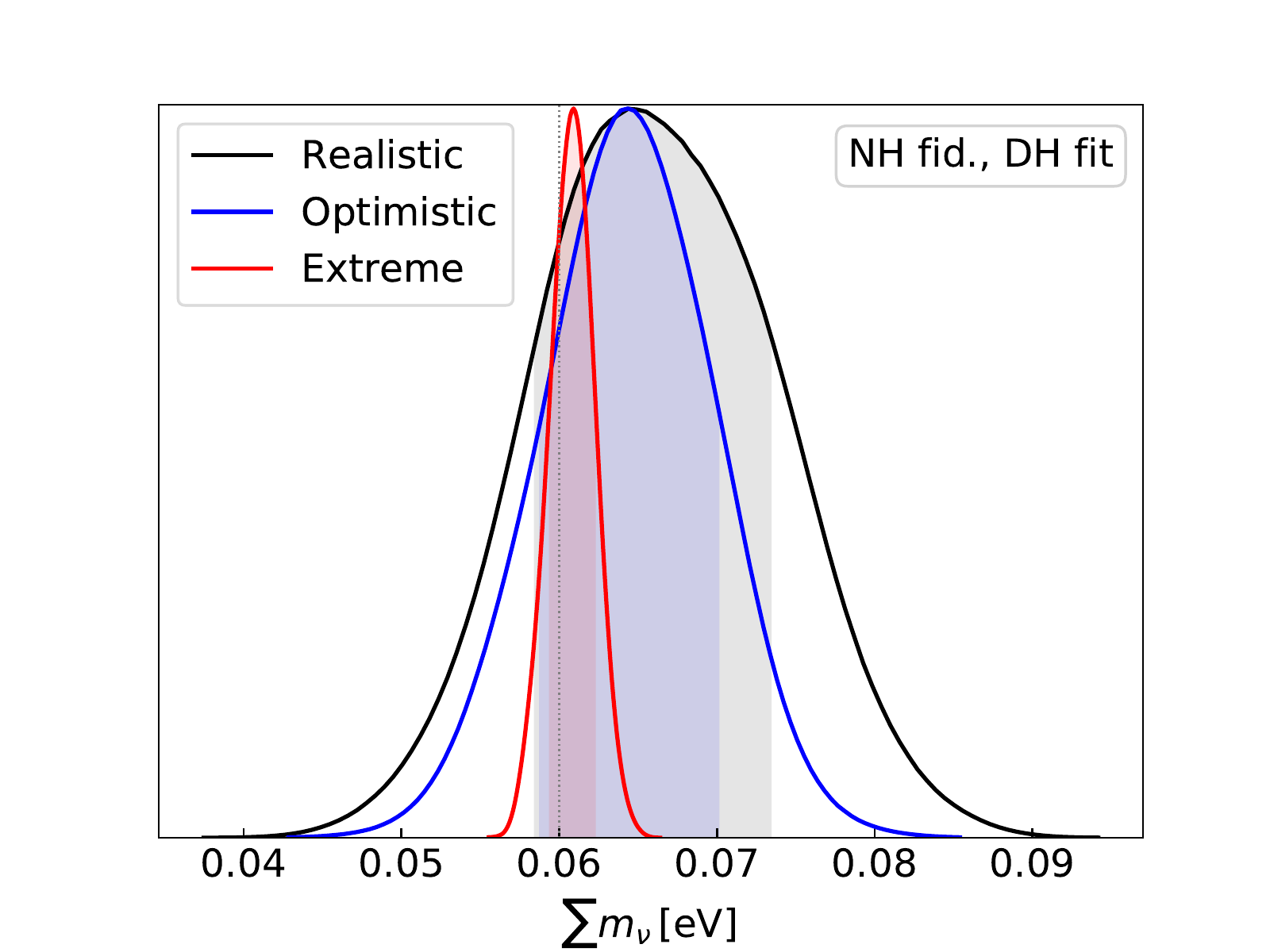}&\includegraphics[width=.5\linewidth]{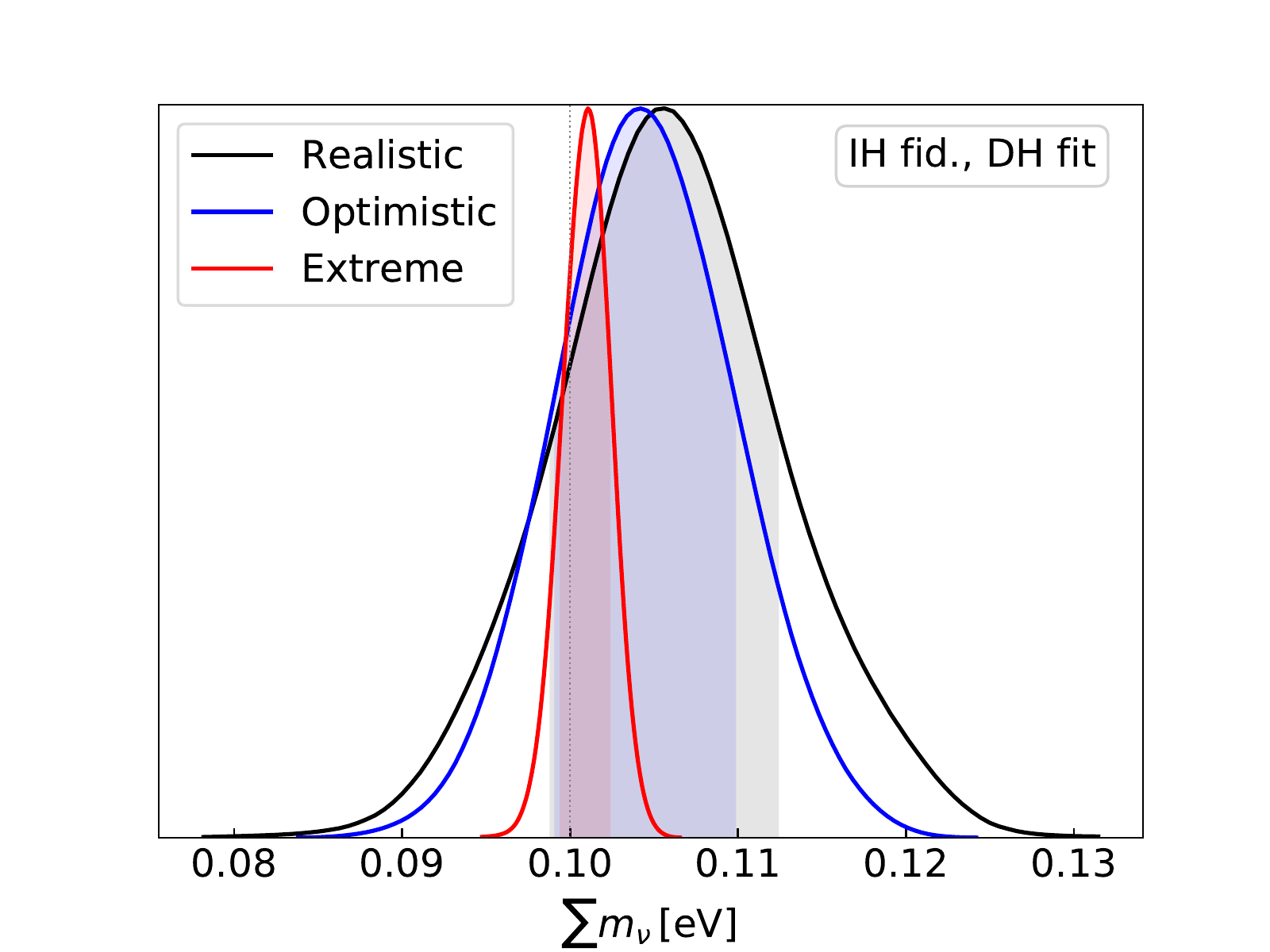}\\
\hfill
\end{tabular} 
\caption{\label{fig:posterior_min} Same as Fig.~\ref{fig:posterior}, but here the fiducial values are set to the minimum mass of each hierarchy, i.e. $\sum m_\nu=0.06$ eV for NH ({\it left panel}) and $\sum m_\nu=0.10$ eV for IH ({\it right panel}).}
\end{figure}

For completeness, we perform the forecast also for the minimum neutrino mass sum in both hierarchies\footnote{For the minimum mass scenario we do not compute the Bayes factor because the different prior volume would affect the result.}; the posterior obtained for $\sum m_\nu$ (Fig.~\ref{fig:posterior_min}) demonstrates that even in this case, with the maximum deviation between NH/IH and DH, the input mass is recovered within $1\,\sigma$, with a systematic bias \cite{Massey:2012cz} of about $\sim 0.5\,\sigma$ in the ``extreme'' scenario, and $\sim 0.7 - 0.8 \, \sigma$ in the ``realistic'' and in the ``optimistic'' scenario.
Thus, we find that in every case and within $1\,\sigma$, the degenerate mass approximation does not affect the estimate of the true value of the neutrino mass sum.
\begin{figure}[tbp]
\centering
\begin{tabular}{cc}
\includegraphics[width=.5\linewidth]{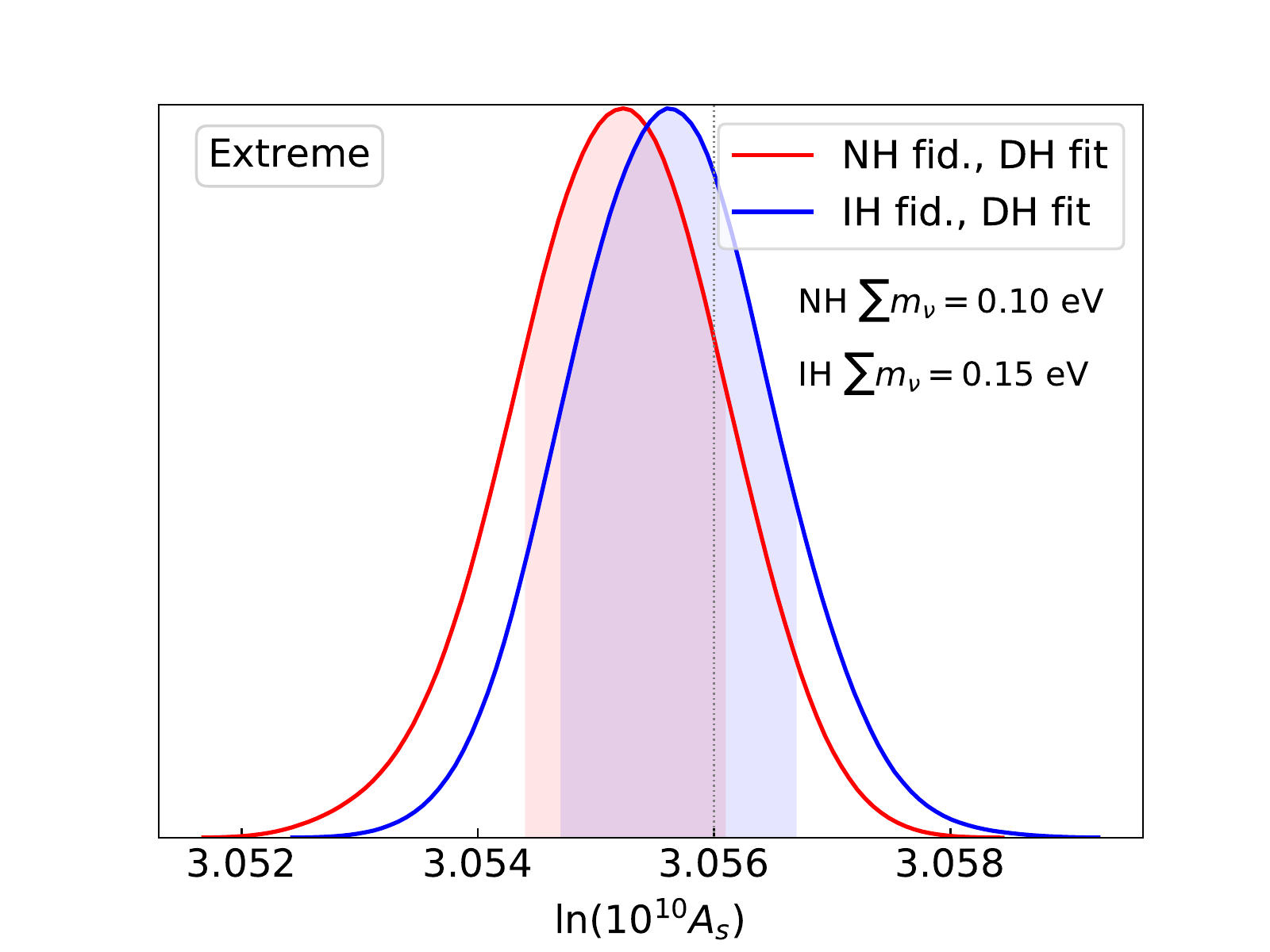}&\includegraphics[width=.5\linewidth]{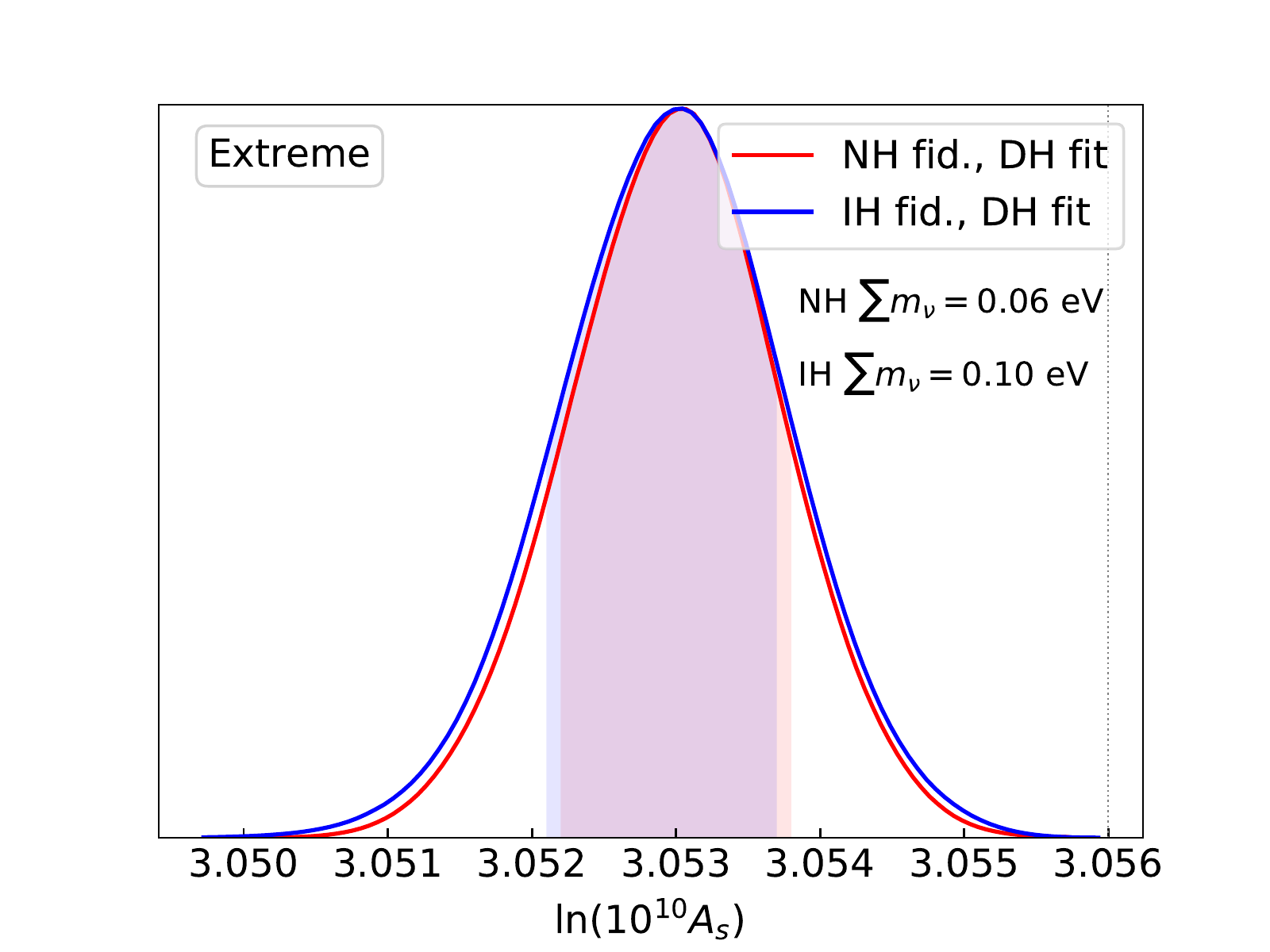}\\
\hfill
\end{tabular}
\caption{\label{fig:posterior_As}Left panel: Marginalized posterior distribution of the amplitude of the primordial power spectrum obtained by fitting the NH with $\sum m_\nu=0.10$ eV (red) or IH with $\sum m_\nu=0.15$ eV (blue) fiducial with the degenerate approximation in the ``extreme''  scenario. Right panel: Same as left panel, but here for the minimum mass of each hierarchy, i.e. $\sum m_\nu=0.06$ eV for NH and $\sum m_\nu=0.10$ eV for IH.}
\end{figure}

The systematic bias affecting the other cosmological parameters is always below $0.5\,\sigma$ ($0.6\,\sigma$) in the ``realistic'' (``optimistic'') case, while
in the ``extreme'' case it is $<0.8\,\sigma$ when considering larger fiducial masses, but it can exceed $1\,\sigma$ in the minimum mass scenario.
In this case the largest bias is found in the estimated mean value of $\ln (10^{10}A_{\rm s})$, which is shifted by more than $3\sigma$ with respect to the fiducial value (see Fig.~\ref{fig:posterior_As}).
The presence of such a systematic bias indicates that, in the very unrealistic hypothesis of a perfect control of systematic effects and theoretical errors, the deviation in the observables is above the observational error (see Section \ref{sec:lss}), and thus can in principle be detected. However, as it often occurs in cosmology, this potential signal is anyway limited by degeneracies with other cosmological parameters that can compensate for it.

\section{Discussion and conclusions}
\label{sec:conclusions}

Over the past decade cosmology has been providing increasingly accurate constraints on neutrino physics.
In the near future, while a significant evidence for a non zero neutrino mass sum seems to be within the capacity of future cosmological surveys, it is still under debate whether it will be possible to disentangle the single mass eigenstates.

In this paper we have discussed in detail the physical effects related to the neutrino mass splittings on the background and perturbation quantities determining the cosmological observables at high redshifts (CMB) as well as in the local Universe (galaxy clustering and weak lensing).
We have quantified this impact in terms of the deviation of each hierarchy (Normal Hierarchy - NH, and Inverted Hierarchy - IH) with respect to the degenerate case (DH), at the minimum mass allowed by oscillations ($0.06$ eV for NH and $0.10$ eV for IH). This choice is motivated by the fact that the deviation of IH or NH from DH are larger than those between IH and NH, as we have shown for both CMB and LSS observables.
For CMB, the lensed and unlensed spectra of CMB temperature, E-modes and B-modes polarization, show deviations at the per mille level, i.e., within cosmic variance.
Concerning low redshift probes, the relative difference in the weak lensing angular power spectrum between NH or IH and DH is $\lesssim 0.4$\%, within the observational error of a future photometric survey like Euclid. On the other hand, the relative difference in the galaxy clustering power spectrum ($\lesssim 0.6$\%), although above the observational error, is well below the modeling uncertainties (e.g., redshift space distortions, scale dependent non-linear bias, non-linear clustering, baryonic feedback) that affect the reconstruction of the observables from raw data.

Besides studying the impact of the neutrino mass splittings at a theoretical level, we have also performed a MCMC forecast of the sensitivity of forthcoming and futuristic cosmological surveys to the hierarchy, computing the Bayesian evidence of NH or IH versus DH.
We obtained that the Bayesian evidence of either IH or NH over DH is inconclusive. 
Indeed, even if we were able to keep under control all theoretical and systematic errors, and to model the galaxy power spectrum at $\sim 0.1$\% accuracy down to very small scales $k>10\,h/{\rm Mpc}$, the presence of correlations with other cosmological parameters would be enough to mimic the tiny deviations induced by the hierarchies.
As a consequence of our theoretical discussion, this result also implies that it is unlikely that cosmology {\it alone} will ever be able to provide a {\it direct} detection of the neutrino mass hierarchy. The only exception might be represented by futuristic 21 cm surveys (e.g., the Fast Fourier Transform Telescope \cite{Tegmark:2008au}) mapping an extremely large volume of the Universe, thus, constraining the growth of structure at the epoch of reionization (or even at the cosmic dawn), when it was not affected by non-linearities or baryonic feedback. However, the presence of foregrounds makes it unclear whether it will ever be possible to reach such an accuracy from 21 cm intensity mapping.
Therefore, a direct detection of the neutrino mass hierarchies relies on future ground-based neutrino oscillation experiments, such as the Deep Underground Neutrino Experiment (DUNE) \cite{Acciarri:2015uup, Ternes:2019sak} and the Jiangmen Underground Neutrino Observatory (JUNO) \cite{An:2015jdp}.

Nevertheless, if nature has chosen the neutrino mass to be (close to) the minimum of the normal hierarchy, cosmology will be able to measure it with such high precision that this will rule out the inverted hierarchy with high statistical significance (see, e.g., Refs.\ \cite{Hamann:2012fe, Brinckmann:2018owf}).
However, even in this best case scenario, fits to purely cosmological data can be carried out assuming the degenerate mass approximation without introducing any bias in the results.

Finally, consider that embedding the prior knowledge on neutrino mass splittings into the cosmological analysis, although physically motivated, might hide the presence of stumbling blocks within the cosmological data, their modeling, or the cosmological model itself.
For instance, if cosmological data prefer a massless neutrino Universe or a neutrino mass sum smaller than the minimum allowed by oscillations, it will be important to observe this discrepancy and to investigate its cause.
A careful comparison between cosmological and laboratory results should always be performed before combining cosmological data with neutrino oscillations in a joint analysis.

\section*{Acknowledgements}
This work was supported by the Villum Foundation. 
MA acknowledges the computing support from INFN-CNAF.

\appendix

\clearpage

\bibliography{references}{}
\bibliographystyle{JHEP}

\end{document}